\documentclass{emulateapj}
\usepackage{natbib} 
\newcommand{\be}{\begin{equation}}
\newcommand{\ee}{\end{equation}}

\begin{document}


\title{Spectro-photometric and weak lensing survey of a supercluster
\\and typical field region -- I. Spectroscopic redshift measurements}

\shorttitle{Spectro-photometric study of a supercluster and field region}

\author{
R.~E.~Smith\altaffilmark{1,3}
H.~Dahle\altaffilmark{2}, 
S.~J.~Maddox\altaffilmark{1}, 
P.~B.~Lilje\altaffilmark{2}
}


\altaffiltext{1}{
School of Physics and astronomy, 
University of Nottingham, 
Nottingham, NG7 2RD, Nottingham}

\altaffiltext{2}{
Institute of Theoretical Astrophysics, 
University of Oslo,
P.O. Box 1029, Blindern, 
N-0315 Oslo, 
Norway}

\altaffiltext{3}{
Department of Physics and Astronomy, 
University of Pennsylvania, 
209 South 33rd Street,
Philadelphia, PA 19104, 
USA.}

\shortauthors{Smith et al.}


\begin{abstract}
  We present results from a spectroscopic study of $\sim$4000 galaxies
  in a $\sim6.2$\,deg$^2$ field in the direction of the Aquarius
  supercluster and a smaller typical field region in Cetus, down to
  $R<19.5$. Galaxy redshifts were measured using the Two Degree Field
  system on the Anglo-Australian Telescope, and form part of our wider
  efforts to conduct a spectro-photometric and weak gravitational
  lensing study of these regions.  At the magnitude limit of the
  survey, we are capable of probing $L_*$ galaxies out to $z\sim0.4$.
  We construct median spectra as a function of various survey
  parameters as a diagnostic of the quality of the sample.  We use the
  redshift data to identify galaxy clusters and groups within the
  survey volume. In the Aquarius region, we find a total of 48
  clusters and groups, of which 26 are previously unknown systems, and
  in Cetus we find 14 clusters and groups, of which 12 are new.  We
  estimate centroid redshifts and velocity dispersions for all these
  systems. In the Aquarius region, we see a superposition of two
  strong superclusters at $z=0.08$ and $z=0.11$, both of which have
  estimated masses and overdensities similar to the Corona Borealis
  supercluster.
\end{abstract}



\section{Introduction}

Over the past few years our understanding of the large scale
distribution of galaxies has advanced greatly. With current
spectroscopic surveys of the local galaxy population ($z\sim0.1$),
such as the 2dF Galaxy Redshift Survey
\citep[2dFGRS;][]{Collessetal2001} approaching 250,000 galaxies, and
the ongoing Sloan Digital Sky Survey \citep[SDSS;][]{Straussetal2002}
set to capture the spectra of a million galaxies, statistical errors
on key quantities of interest are fast becoming diminutive. A major
goal of these surveys is to make accurate measurements of the
large-scale distribution of \emph{matter} density fluctuations in the
Universe and so measure the cosmological parameters.

However, since one is not probing the matter directly, but instead
some population of objects that merely trace the mass, e.g., galaxies
of a particular luminosity class, spectral type etc..., it is
necessary to make some assumptions about the way in which these two
distributions are related -- or more commonly biased. In recent work
\citep{Tegmarketal2004,Percivaletal2004}, it is assumed that there is
a certain large scale, $\sim20 h^{-1}$Mpc or so, above which the
galaxy tracers and mass are related by a simple scale independent
constant bias.  Whilst this may indeed turn out to be correct, it is
important to establish the veracity of this assumption through direct
observation, if redshift surveys are to be trusted as precision
cosmological probes.

Weak gravitational lensing provides the most exciting progress towards
solving this problem. The weak distortions in the shapes of faint
background galaxies due to the gravitational deflection of light
bundles as they travel through distorted spacetime provides a direct
probe of the projected mass distribution
\citep[e.g.,][]{Kaiser1992,KaiserSquires1993}. More recently
\citet{BaconTaylor2003} and \citet{Tayloretal2004} have shown that if
one has redshift information for all galaxies, then one may also
reconstruct the 3D mass distribution.

Also, \citet{Schneider1998} has developed a method for exploring the
bias. This has been applied by
\citet{Hoekstraetal2001,Hoekstraetal2002}, who found on scales
$<5h^{-1}$Mpc, that the bias was scale-dependent, with positive bias
on scales $<0.25h^{-1}$Mpc, anti-bias on scales $\sim 1h^{-1}$Mpc and
almost no biasing at the limits of their survey.  If all of the
systematics in this work are under control, then this is strong
support for the `Concordance cosmology' \citep{Wangetal2000}, which
requires bias of this kind to match the galaxy clustering statistics
\citep{Bensonetal2000,PeacockSmith2000}.

A complicating factor in the analysis is that the lensing effect
depends on the geometry of the system and one therefore requires
knowledge of the spatial distribution of the lens and source galaxy
populations to obtain accurate measurements.  However, as
\citet{Hoekstraetal2001,Hoekstraetal2002} have shown, if one has
redshift information for the lens galaxies only, then one may make
good progress through assuming an appropriate redshift distribution
for the background lens galaxies.

In a series of papers, we perform a combined weak lensing and
spectro-photometric study of a supercluster and also a control field
region with the specific aim of uncovering detailed information about
bias over a wide range of scales and environments.  We intend to
bridge the gap between having no and full spatial information along
the light cone, by measuring spectra for the foreground galaxies. In
this, the first paper of the series, we present the spectroscopic part
of the survey.  In a subsequent paper we will describe our deep
photometric and weak lensing analysis of the regions, and in a further
paper we will combine all of the data to directly answer the above
questions.

Using the 2dF instrument on the AAT we have measured the spectra for a
significant fraction of all galaxies in the survey regions, down to
$R<19.5$. From these data we determine redshifts through
cross-correlating each spectrum with a set of standard galaxy and
stellar templates. We then determine the redshift space distributions
for the two survey regions, finding marked differences between them.
As a by-product of this work, we have used the full redshift space
sample to compile new lists of cluster candidates for the regions, and
this is the subject of the latter part of the paper. For these cluster
candidates we estimate redshifts and velocity dispersions.

The paper is broken down as follows: In \S~\ref{sec:population}, we
describe the survey and the spectroscopic target catalogs. In
\S~\ref{sec:spectroscopy}, we describe the observations and data
reduction.  In \S~\ref{sec:properties}, we describe the redshift
estimation procedure and then present some interesting properties of
the spectroscopic sample. In \S~\ref{sec:cluster} we use the sample to
establish clusters within the survey. Here, we also provide an updated
cluster catalog for the Aquarius region. In \S~\ref{sec:velocity} we
give estimates of the centroid redshift, velocity dispersion and
asymmetry index of each cluster candidate. Finally, in
\S~\ref{sec:summary}, we discuss our findings and summarize the
results.  Throughout this paper we have assumed concordance cosmology
\citep{Wangetal2000} with $\Omega_m=0.3$, $\Omega_{\Lambda}=0.7$,
where $\Omega_m$ and $\Omega_{\Lambda}$ represent the matter and
vacuum energy density parameters at the present epoch. All celestial
coordinates are given for epoch J2000.0.


\begin{figure*}[t]
  \centering{\includegraphics[width=8cm,angle=-90,clip=]{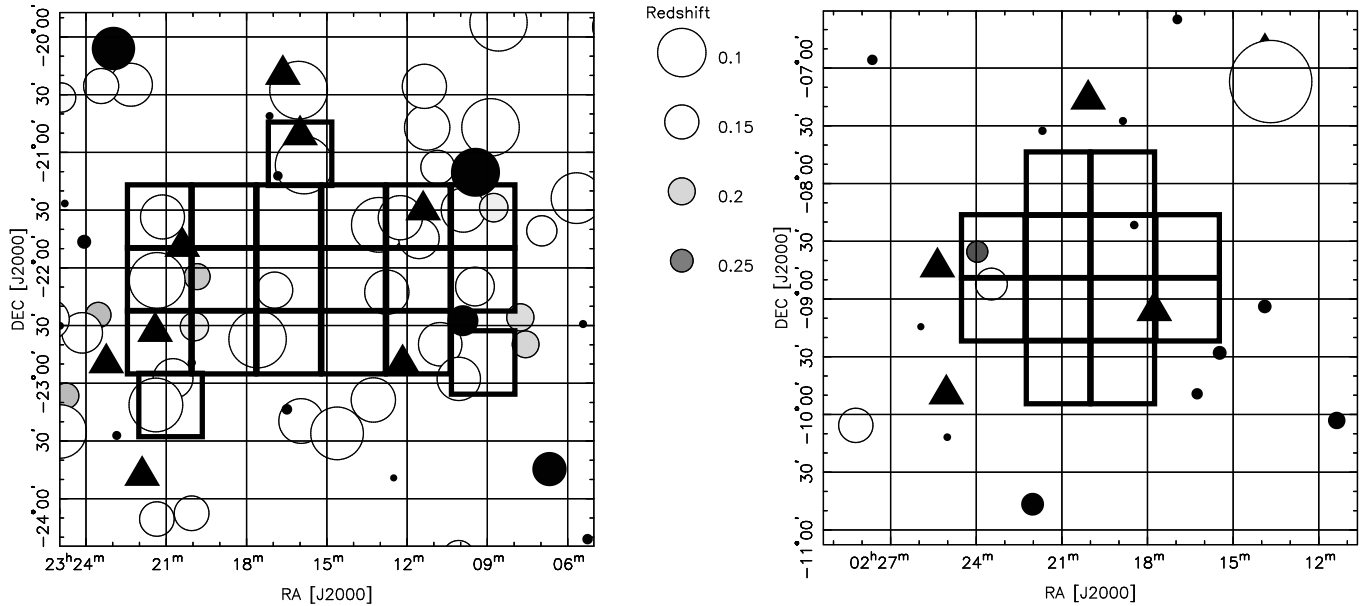}}
  \caption{\small{Survey regions. The left panel shows the Aquarius
      supercluster region and the right panel shows the typical field
      region in Cetus. In both panels the contiguous thick square
      tiles correspond to the field of view of the WFI camera
      ($33'\times34'$), and de-mark the survey boundaries. Overlaid on
      both plots are known clusters; those that have had redshifts
      measured are represented by circles, where the size and shading
      of the circle is in proportion to the Abell radius and redshift
      of the cluster; and those with no previous redshift estimate are
      identified as triangles. Filled black circles represent field
      stars brighter than $V=8.0$, where the size of the circle
      increases with increasing brightness.}
    \label{fig:WFIfields}}
\end{figure*}


\section{Survey Description}\label{sec:population}

The survey comprises deep photometric imaging and spectroscopy in two
regions: the supercluster in the direction of Aquarius
[$\alpha=23^h15^m$, $\delta=-22^{\circ}$], which is of general
interest; and a typical field region in Cetus [$\alpha=02^h19^m,
\delta=-8^{\circ}50'$]; hereafter referred to as the Aquarius and
Cetus regions. The Cetus region was chosen to coincide with one of the
SDSS southern strips. In the future this will provide us with
additional multi-colour photometry and spectra. The deep photometric
observations are being carried out on the ESO 2.2m telescope at La
Silla, using the Wide Field Imaging Camera (WFI) and the spectroscopy
is being carried out using the 2dF multi-fiber-object spectrograph on
AAT.

The Aquarius supercluster was firmly established by
\citet{Batuskietal1999}, previously being noted as one of the highest
concentrations of rich clusters on the sky \citep[~hereafter
ACO]{Abell1961,Murrayetal1978,Ciardulloetal1985,Abelletal1989}.  From
spectroscopy, \citeauthor{Batuskietal1999} deduced that the core of
the supercluster contained five richness class $R \ge 1$ ACO clusters:
\objectname{A2546}, \objectname{A2554}, \objectname{A2555},
\objectname{A2579} and \objectname{A3996}; making the Aquarius
supercluster even denser than the most massive superclusters at lower
redshifts, such as the Shapley supercluster \citep{Raychaudhury1989}
and the Corona Borealis supercluster \citep{Postmanetal1986}. More
recently a study of the region was undertaken by \citet[~hereafter
Caretta02]{Carettaetal2002}, who used projected galaxy distributions
and follow-up spectroscopy to compile a list of 102 cluster candidates
over $\sim$80~deg$^2$.  There also exists archival Chandra and/or ASCA
data for 10 of the Aquarius clusters.

Fig \ref{fig:WFIfields} presents the survey regions.  The Aquarius
field comprises 20 overlapping WFI pointings, covering an angular area
of $\sim$6.2 deg$^2$, which at the median redshift of the
spectroscopic survey, ($\bar{z}\sim 0.176$), corresponds to scales of
the order 30 $h^{-1}$ Mpc $\times$ 15 $h^{-1}$ Mpc. Out to the
effective limit of the survey $(z\sim0.45)$, the comoving volume
surveyed in the Aquarius region is $1.0\times10^6h^{-3}$Mpc$^3$. The
Cetus field comprises 12 overlapping WFI pointings, covering an
angular area of $\sim$3.7 deg$^2$, which again at the median redshift
of the spectroscopic survey corresponds to transverse scales 20
$h^{-1} $Mpc $\times$ 20 $h^{-1}$ Mpc; and out to the effective survey
limit, a comoving volume of 5.6 $\times10^5\,h^{-3}$Mpc$^3$.

Within the Aquarius fields we have identified 23 known clusters
(Caretta02), 5 of which have no redshift estimate. This corresponds to
an average surface density of 3.7 clusters deg$^{-2}$. In contrast,
the Cetus region was chosen to contain only one Abell cluster,
which had a richness class $R=0$ and distance class $D=6$. This gave an
average 0.27 clusters deg$^{-2}$. Since the mean surface density of
Abell clusters over the whole sky is $\sim$0.2 clusters deg$^{-2}$, we
are justified in our assertion that the Cetus region is indeed
a `typical' patch of sky. We note that there is an additional known
cluster, \objectname{RX J0223.4-0852}, within the field, and 
an Abell cluster, \objectname{A348} ($z=0.27$), that glances the 
Eastern edge of our 2dF survey region.


\subsection{The Spectroscopic Input Catalog}

The spectroscopic targets were taken from the SuperCOSMOS\footnote{The
  on-line SuperCOSMOS archives can be found at: {\ttfamily
    http://www-wfau.roe.ac.uk/sss/}} on-line catalog
\citep{Hamblyetal2001a}. We selected all objects down to a magnitude
limit of $R\le19.5$. For $R\le19$, the image classification is $>90\%$
reliable and for $R\le19.5$ this drops to $>80\%$
\citep{Hamblyetal2001b}. On inspecting the data we found that faint
satellite trails still remained in the catalog. These were removed by
matching objects from the $R$ and $b_J$ plates, which were observed at
separate epochs.  The astrometry of the SuperCOSMOS R-band data has a
measured positional accuracy to better than $\pm 0.2\arcsec$ at
$R\sim18$, deteriorating to $\pm 0.3\arcsec$ at $R\sim21$
\citep{Hamblyetal2001b}, and this was of sufficient accuracy for the
requirements (2$\arcsec$ fiber diameters) of the 2dF instrument.  We
then corrected the $R$ magnitudes for absorption using the dust maps
of \citet{Schlegeletal1998}. For the Aquarius region we used an
extinction correction $A_R\sim0.074$, and for Cetus we used
$A_R\sim0.07$. 

Taking $M_{*r}-5\log_{10}h=-20.83$ \citep{Blantonetal2003}, then for
galaxies at the flux limit of our survey ($R=19.5$) we are capable of
probing $M_{*}$ galaxies out to $z\sim0.4$, or equaivalently angular
diameter distances of $\sim1$Gpc, where we have assumed a
$K$-correction $\sim0.5$ \citep{Colemanetal1980}.

\begin{figure}[t]
\centering{\includegraphics[width=8.5cm,angle=0,clip=]{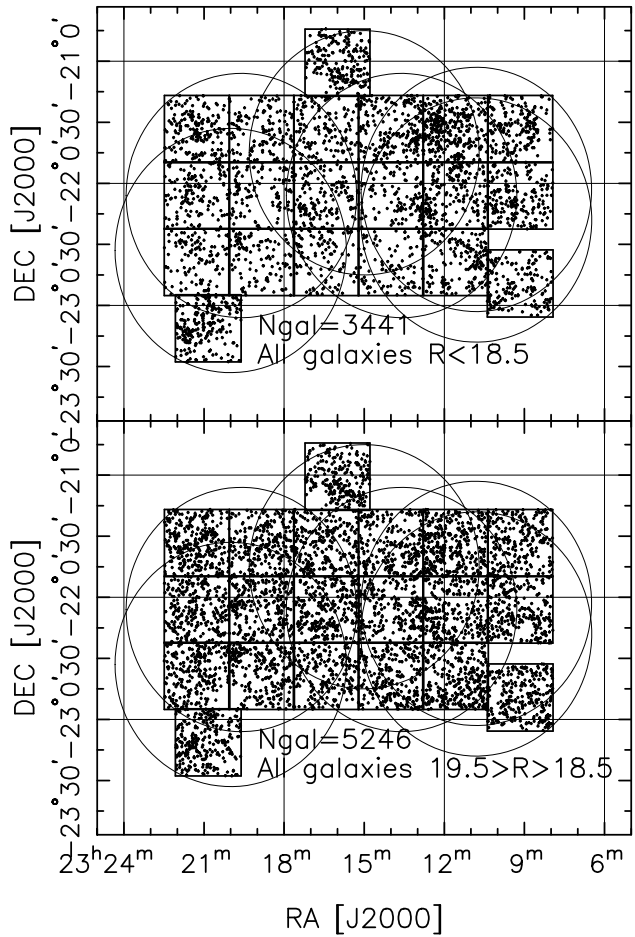}}
\caption{\small{Source catalogs for the survey. Top panel shows
    all galaxies in the survey region with ($R<18.5$) and bottom panel
    shows all galaxies with ($19.5>R\ge18.5$). Overlaid thick grid
    represents the fields that have been selected for deep $R$ band
    imaging using the WFI camera. Also overlaid are the 2dF fields
    (large circles).}
\label{fig:SourceCat}}
\end{figure}

To acquire spectra of sufficient quality for precise redshift
estimation, we followed the 2dFGRS strategy and aimed to obtain
spectra with a minimum signal-to-noise ($S/N$) $\sim$10. For galaxies
with a $S/N$ of this order, \citet{Collessetal2001} found a redshift
completeness $>90\%$. For objects at the survey flux limit and and in
poor seeing conditions, we estimated that these $S/N$ levels could be
reached for 6500~s integrations. However, for galaxies one magnitude
brighter, a similar $S/N$ was achieved in roughly half the time. To
optimize the use of telescope time, the spectroscopic observations
were split into two sets: short exposures for bright ($R<18.5$)
galaxies; and longer exposures for faint ($19.5<R\le18.5$) galaxies.
Fig.\ref{fig:SourceCat}, top panel, presents the input catalog for the
bright Aquarius galaxies and the bottom panel shows the faint. For the
Aquarius catalogs we identified $3441$ bright galaxies and $5246$
faint. Similarly, for the Cetus region catalog we identified $1414$
bright galaxies and $1881$ faint.


\section{Spectroscopic Observations and Analysis}\label{sec:spectroscopy}


\subsection{Fiber Configurations}\label{ssec:fibconfig}

For the Aquarius field, six overlapping 2dF pointings were chosen to
cover the target area, and for the Cetus field, the whole target area
was contained within a single 2dF centre. Owing to the high surface
density of sources, $\sim$1400 deg$^{-2}$ in the Aquarius region and
$\sim$1048 deg$^{-2}$ in the Cetus region, multiple 2dF pointings of
each of the chosen centres were required to target all of the galaxies
in each field. In full operational mode the 2dF is capable of
acquiring the spectra of 400 objects simultaneously. The fiber
arrangements for individual fields were pre-configured using the
observatory supplied software package {\small{\ttfamily CONFIGURE}}.
For a full description of the 2dF instrument and configuring for dense
fields see \citet[hereafter BGB02]{Baileyetal2002}.


For the bright galaxy fields we typically allocated 360 fibers to
target objects and 40 to sky (20 per spectrograph). For the faint
galaxy fields we increased the allocation of sky fibers to 30 per
spectrograph. This increase was to ensure accurate sky subtraction for
the faint spectra. We discuss this further in \S~\ref{ssec:skysub}.
Also as part of our allocation strategy, there was a small chance
that a few objects would be re-observed.  As we describe later in
\S~\ref{ssec:redshifts}, this small number of repeat observations
enabled a useful test of the redshift estimation procedure.


\subsection{Observing}

Observations were carried out between October 29 and November 2, 2002
at the AAT. The 2dF spectrographs were configured with the 316R and
270R gratings. This pair gave FWHM resolutions of 8.5 and 9.0 \AA \
per pixel, respectively, and an efficiency $>70\%$ for the wavelength
range 5000-8500 \AA. For the bright catalog galaxies, individual
pointings comprised: calibration frames, consisting of a tungsten lamp
flat field exposure, followed by a CuHe+CuAr arc lamp exposure;
$3\times 1000$~s exposures of the target field. The target spectra
were acquired as a series of snapshots to facilitate the removal of
cosmic rays. For the faint catalog galaxies, the procedure was
identical except that 5$\times$1300~s exposures were taken and also a
further arc lamp calibration was taken at the end of the observation.
In total, we acquired 6800 target spectra.



\subsection{2dF Data Reduction}\label{ssec:skysub}

All data reduction was performed using the latest version of the
AAO-supplied software package {\ttfamily 2dFDR-V2.3} (for details, see
BGB02).  In brief, the reduction recipe that we adopted was as
follows: The data were de-biased. Bias strips were trimmed from the
data frame and the mean of these was subtracted from the data. The
fiber-flat field frame was then used to generate tram-line maps for
the data. Since the accurate determination of these was crucial to
achieving reliable reduction, they were carefully inspected at each
stage in their generation.

\begin{figure}[t]
  \centering{
    \includegraphics[width=9.5cm,angle=-90,clip=]{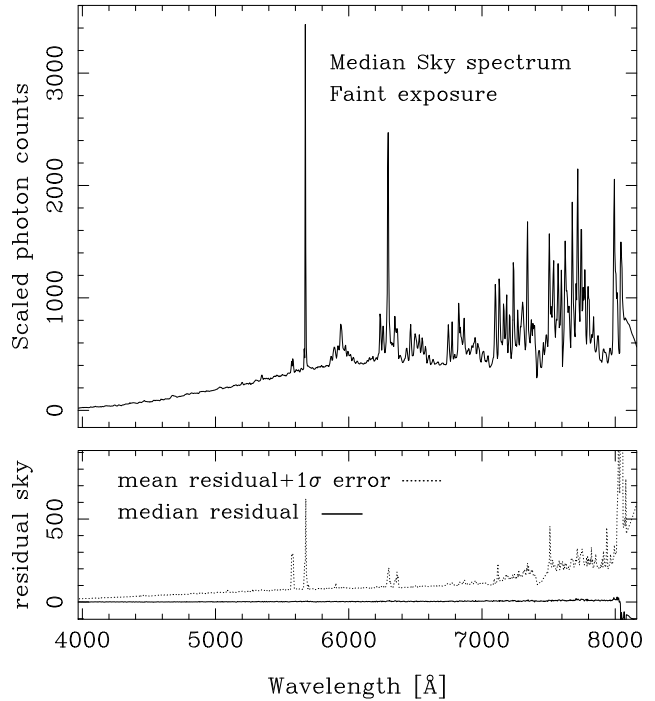}}
  \caption{\small{Top panel shows the median sky spectrum. Bottom panel
      shows the median residual sky (solid line) and the
      mean$+1\sigma$ residual sky (dotted line), determined from all
      of the sky fibers in the deep exposure frames.}
    \label{fig:MedSky}}
\end{figure}

The spectra were then extracted using the {\ttfamily FIT} routine,
which performs an optimal extraction based on fitting overlapping
Gaussian profiles to a fiber flat-field frame.
%
The extracted data were then flat fielded using the fiber flat field
data and re-binned onto a uniform linear wavelength scale, with the
wavelength calibration for each fiber being determined from the
closest in time CuHe+CuAr arc lamp exposure. The relative throughput
of each fiber was then determined and the median sky generated and
subtracted from each spectrum.  Since good determination of the
relative fiber throughput and sky subtraction was crucial for the
faint galaxy spectra, we considered this carefully. We found that
through using the {\ttfamily Skyflux(cor)} routine in the 2dF
reduction package, we were able to produce good throughput
calibration, and perform sky subtractions to an accuracy
$\sim1.5-3.0\%$. These accuracies are comparable to those obtained by
\citet{Willisetal2001}.  In Fig. \ref{fig:MedSky}, we show the median
sky and residual sky for all of the faint galaxy exposures. The
residual sky was constructed by subtracting the median sky spectrum
from each of the individual sky fibres and then calculating the median
spectrum of these sky-subtracted sky fibres. This clearly demonstrates
that the sky subtraction method that we have adopted is accurate. For
the bright data the overall sky subtraction was of similar quality.
Finally, the reduced runs were optimally combined and cosmic ray
events were removed using a $5\sigma$-clipping rejection criterion.


\begin{figure}
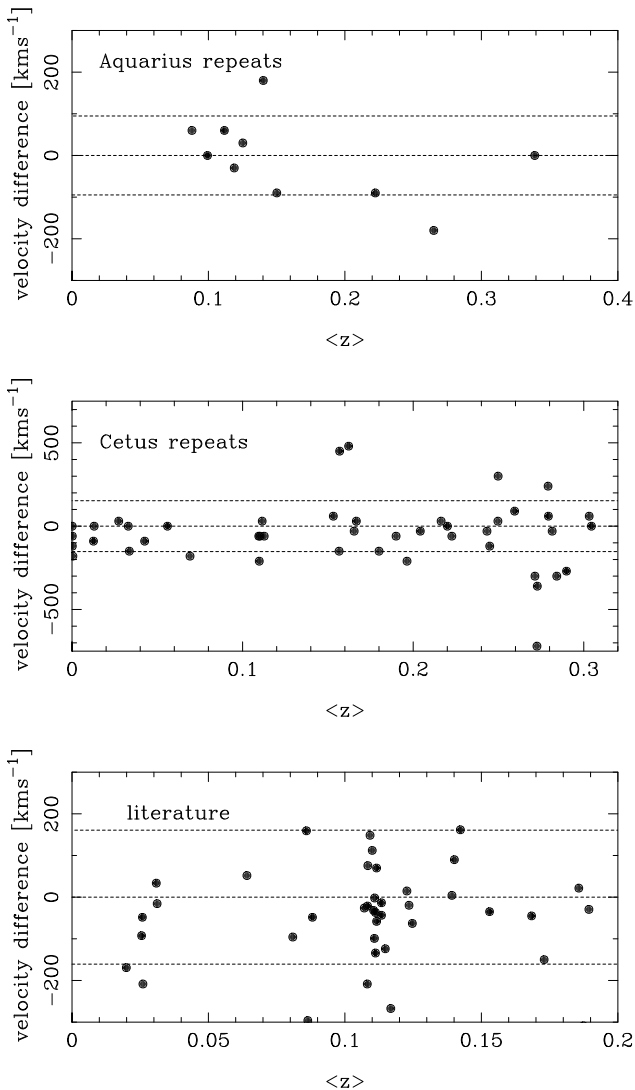

  \centering{\includegraphics[width=4.9cm,angle=-90,clip=]{f4a.eps}}
  \centering{\includegraphics[width=4.9cm,angle=-90,clip=]{f4b.eps}}
  \centering{\includegraphics[width=4.9cm,angle=-90,clip=]{f4c.eps}}
  \caption{\small{Comparison of redshifts for targets with repeated
      observations in the Aquarius region (top panel) and the Cetus
      region (middle panel). The bottom panel shows targets with
      previously measured redshifts. Plotted are the velocity
      difference $\Delta(cz)$ as a function of the mean redshift of
      the two observations $<z>$. In all panels, the points represent
      spectra for which both of the estimated qualities were, $Q\ge3$.
      The dash lines represents the {\em rms} variation.}
    \label{fig:Duplicate1}}
\end{figure}


\begin{figure*}
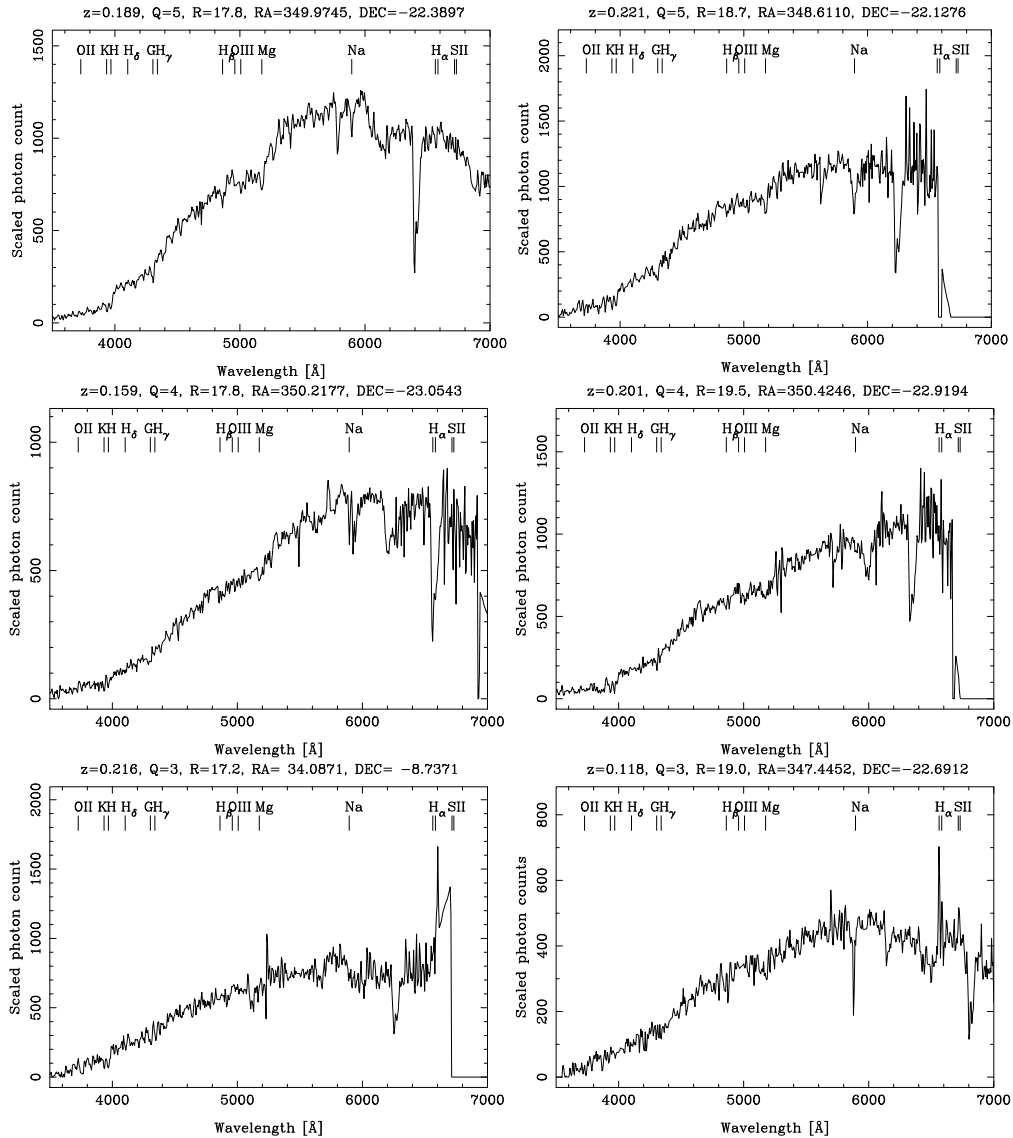

  \centering{
    \includegraphics[width=5.cm,angle=-90,clip=]{f5a.eps}
    \includegraphics[width=5.cm,angle=-90,clip=]{f5b.eps}}\\
  \centering{
    \includegraphics[width=5.cm,angle=-90,clip=]{f5c.eps}
    \includegraphics[width=5.cm,angle=-90,clip=]{f5d.eps}}\\
  \centering{
    \includegraphics[width=5.cm,angle=-90,clip=]{f5e.eps}
    \includegraphics[width=5.cm,angle=-90,clip=]{f5f.eps}}\\
  \caption{\small{Selection of rest frame spectra. The panels on the 
      left show spectra from the bright ($R<18.5$) survey and those on
      the right show spectra from the faint ($19.5>R\ge18.5$) with
      redshifts ($0.15<z<0.25$). Going from top to bottom shows
      changes in quality, with top panels representing $Q=5$, middle
      $Q=4$ and bottom $Q=3$.  The short thin lines at the top of each
      figure denote interesting absorption/emission features.
    }\label{fig:SelecSpec}}
\end{figure*}


\subsection{Redshift Estimation}\label{ssec:redshifts}

Redshifts were measured using a modified version of the {\ttfamily
  CRCOR} code developed for the 2dFGRS \citep{Collessetal2001}.  This
uses both cross-correlation against template spectra
\citep{TonryDavis1979} and identification of emission lines to
estimate the redshift for each galaxy. The set of templates used for
this analysis was the same as presented by \citet{Collessetal2001}.
Each redshift estimate was visually checked against the galaxy
spectrum, and a quality flag, $Q$, was assigned according to the
reliability of the estimate. This reliability assessment was arrived
at through considering the cross-correlation $R$ value and the number
of identified emission lines. The value of $Q$ assigned to each
estimate was between 1-5, with 1 being no confidence and 5 being
extremely confident. In all our later analysis we use only galaxies
with $Q\ge 3$.  Of the 5875 target objects observed, 3841 had $Q\ge3$.
Of these 1245 were $Q=3$, 2251 were $Q=4$ and 345 were $Q=5$. Note
that poor weather observations, for which only a handful of spectra
were measurable, have also been included in this accounting.

During the observing run, 15 galaxies in the Aquarius field and 121
galaxies in the Cetus field were targeted twice. These repeats
provided a useful test of the redshift estimation process.  In Fig.
\ref{fig:Duplicate1}, we plot the difference in recession velocity,
$\Delta(cz)=c(z_1-z_2)$, between the two estimated redshifts as a
function of the mean redshift $\left<z\right>$ of the pair. For the
Aquarius field (Fig.~\ref{fig:Duplicate1}, top panel), of the 15
repeats, 10 of the redshift estimates agreed very well with an
\emph{rms} error of 67\, km\, s$^{-1}$. For four of the cases one of
the observations was of poor quality $Q<3$, so that the comparison was
unreliable and we do not show these. However, for one of the repeats,
the estimates disagreed by $\Delta (cz)\sim15,000 \,{\rm km\,
  s^{-1}}$, and the quality flags were 3 and 4. We then inspected this
erroneous pair to ascertain whether or not it was a genuine `blunder'.

Inspecting the images from the SuperCOSMOS archive we found two
sources separated by less than 2$\arcsec$. Owing to the strong
likelihood that the spectrum is composite, we were unable to trust the
spectral features. For the Cetus field region
(Fig.~\ref{fig:Duplicate1} middle panel), of the 121 repeats, 70 of
the pairs had at least 1 galaxy with $Q<3$. These we did not consider
further. Of the remaining 51, we found 11 pairs with a velocity
difference greater than 300\, km\, s$^{-1}$.  Inspecting these closer,
we found evidence for 3 having close companions.  Excluding these from
the sample we then found an {\it rms } error of 153\, km\, s$^{-1}$.

A further test of the redshift estimation procedure was performed by
comparing our sample with overlap galaxy redshifts extracted from the
NED \footnote{NED: NASA/IPAC Extragalactic Database, can be found at:
  {\ttfamily http://www.ipac.caltech.edu}}. From this archive we found
42 literature redshifts that overlapped with our Aquarius survey
region. No redshifts were available for the Cetus region.
Fig.~\ref{fig:Duplicate1} (bottom panel) shows the velocity
differences from this comparison. For these data we initially found an
{\it rms } error of $\sim4000$\, km\, s$^{-1}$.  However, on
identifying and removing a single observation for which
$\Delta(cz)=26,400$\, km\, s$^{-1}$, we found that the {\it rms }
dropped to $\sim160$\, km\, s$^{-1}$. For this NED galaxy, we traced
the redshift measurement to the work of \citet{CollessHewett1987}, who
reported that the estimate for this object was not safe and was
rejected from their analysis. This leads us to conclude that our
overall blunder rate for measuring redshifts is low.


\section{Characterizing the Sample}\label{sec:properties} 

\subsection{The Rest-frame Spectra}

Fig.~\ref{fig:SelecSpec} presents a selection of `early-type' rest
frame galaxy spectra with redshifts around the median for the survey,
$(0.15<z<0.25)$. These were generated by linearly interpolating over
important sky absorption bands, de-redshifting each spectrum and then
linearly interpolating it on to a uniform wavelength scale with range
3500\AA$<\lambda<$7000\AA\ and pixel scale 5.83\AA \ per pixel.


\begin{figure}
\centering{\includegraphics[width=8.5cm,angle=0,clip=]{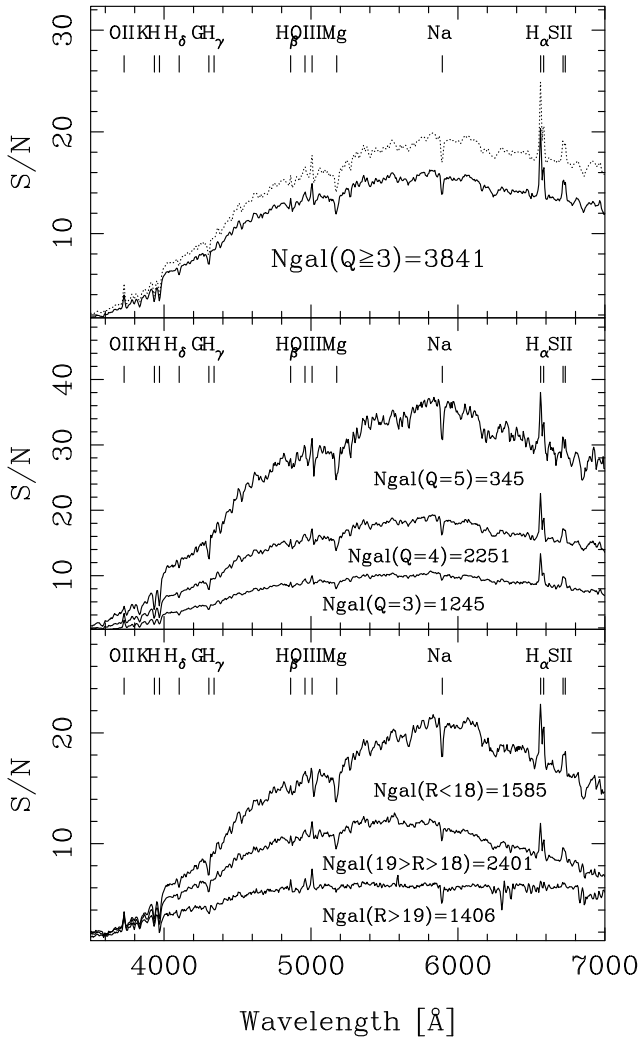}}
\caption{\small{The mean and median $S/N$ rest-frame galaxy
    spectra of the sample. The top panel presents the median (solid
    line) and mean (dotted line) galaxy spectrum for all of the
    galaxies with $Q\ge3$. The middle panel presents the median galaxy
    spectra for the three different quality classes. The bottom panel
    presents the median galaxy spectra that result when the galaxies
    are divided into bins in magnitude: $19.5 \ge R > 19$; $19 \ge R >
    18$; and $18\ge R$.}
\label{fig:MedianAll}}
\end{figure}


\begin{figure}
\centering{\includegraphics[width=8.5cm,angle=0,clip=]{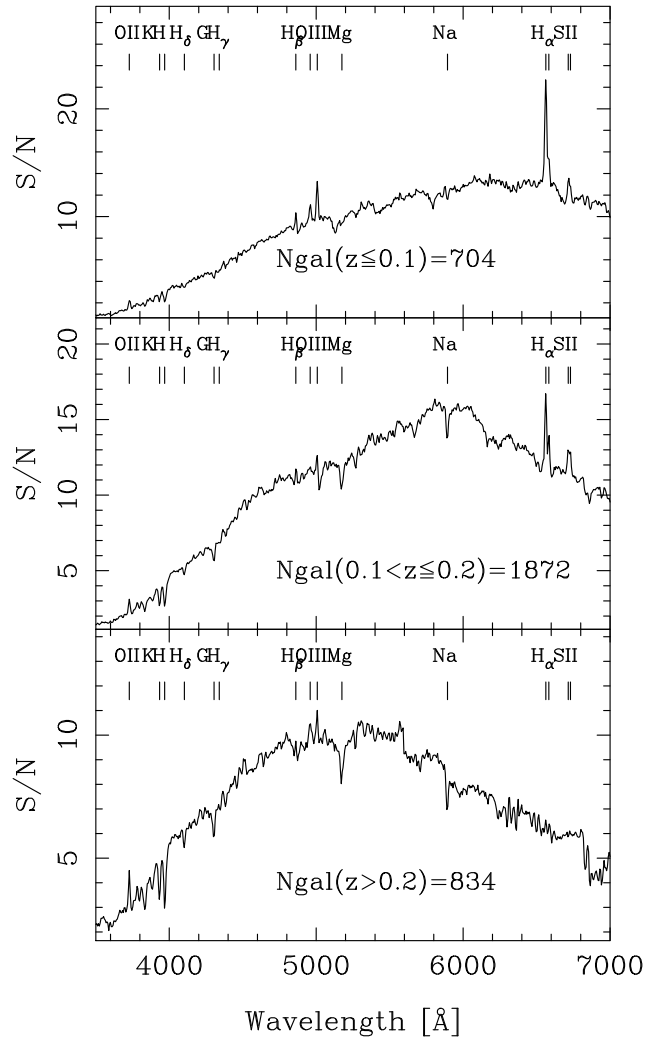}}
\caption{\small{The median rest-frame spectra by redshift. The galaxy
    sample has been split into three bins in redshift space these are:
    all galaxies $z\le0.1$ (top panel); all galaxies with
    $0.1<z\le0.2$ (middle panel); all galaxies $z>0.2$ (bottom
    panel).}
\label{fig:MedianZs}}
\end{figure}


We now calculate the mean and median signal-to-noise rest-frame
spectrum as functions of quality class, magnitude and redshift as
diagnostics for the sample.  Fig.~\ref{fig:MedianAll}, top panel,
shows the mean and median $S/N$ spectrum for galaxies with $Q\ge 3$.
For both mean and median, the general shape of the spectrum is close
to that of an early-type galaxy, but with emission lines superposed.
Furthermore, the average galaxy spectrum in the sample has a $S/N$
better than 10.

Fig.~\ref{fig:MedianAll}, middle panel, shows the median spectrum by
quality class. It is apparent that nearly all of the features in the
high quality spectra are discernible in the lower quality spectra, but
with lower strength absorption and emission features. Quantitatively,
the $Q=3$ and $Q=4$ spectra possess a $S/N$ level a factor of $\sim4$
and $\sim2$ lower than the $Q=5$ spectra.

Fig. \ref{fig:MedianAll}, bottom panel, shows the result of splitting
the sample into three bins in magnitude: $R\ge19.0$; $19.0>R\ge18.0$;
and $R<18.0$. For the brightest two bins, the median spectral energy
distributions are better than $S/N=10$, but for the fainter bin
galaxies the $S/N$ is lower, being $\sim6$. This indicates that these
objects are only just being detected. However, there is clear evidence
of characteristic emission and absorption features, which indicates
that for most cases, the correct redshift has been assigned to these
galaxies.  The absence of H$_{\alpha}$ emission can be attributed to
two effects: The first is that the faint objects are predominantly at
higher redshifts and so have had the red part of their spectrum
redshifted out of the frequency bandwidth for our observations; the
second is that since our sample contains a large number of
low-redshift clusters we are preferentially biased to observing
low-luminosity cluster ellipticals.

Finally, in Fig. \ref{fig:MedianZs}, we consider the median by
redshift. We have split the sample into three bins in redshift:
$z\le0.1$; $0.1<z\le0.2$; $z>0.2$. Considering the first redshift bin,
$z\le0.1$, the median spectrum is of high $S/N$, is flatter than a
typical early-type spectrum, and has weak absorption features and
fairly strong emission lines. The median spectrum for the next
redshift bin, $0.1<z\le0.2$, again is of good $S/N$, is similar to an
early type spectrum but with emission lines superposed. For the
highest redshift bin, $z>0.2$, we find that the median spectrum shows
strong characteristics of being an early-type galaxy, but with strong
\ion{O}{2} and \ion{O}{3} emission features.  Again, the absence of
H$_{\alpha}$ emission can be understood from the arguments discussed
above. It is possible that some of this evolution of the median
spectrum with redshift may be due to true evolution in the galaxy
populations. However, it is more likely due to selection effects.
Since the Aquarius fields are cluster rich, we should expect a large
number of cluster ellipticals in the intermediate redshift bins,
whereas for the low redshift bins, we expect a more even mix of early-
and late-type galaxies. Owing to the flux limit, the higher redshift
bin will be dominated by bright ellipticals and also bright spirals,
which accounts for the strong emission line and absorption features.


\begin{figure}
\centering{\includegraphics[width=4.8cm,angle=0,clip=]{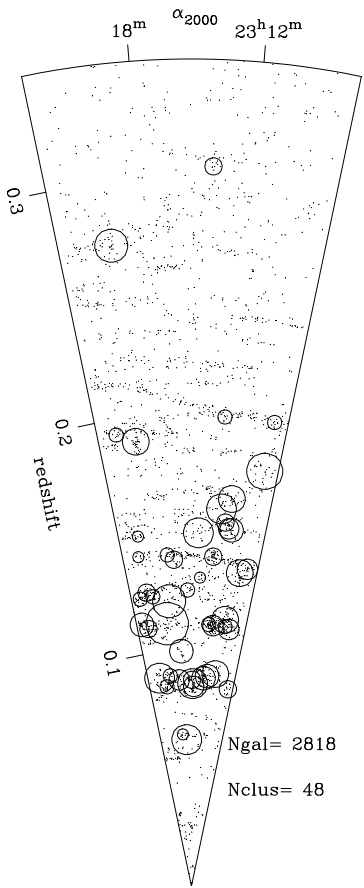}}
\caption{\small{Cone plot for Aquarius region. This shows all 
    galaxies with $Q\ge3$. Circles indicate cluster candidates from \S
    5, where sizes are scaled by the estimated velocity dispersion
    (see Table \ref{tab:AQRlocation}).}
\label{fig:Cone1}}
\end{figure}


\begin{figure}
  \centering{\includegraphics[width=4.8cm,angle=0,clip=]{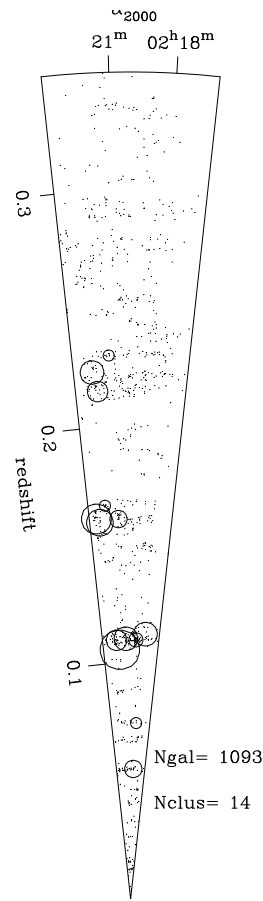}}
  \caption{\small{Similar to figure \ref{fig:Cone1}, but for
      the Cetus region.}
    \label{fig:Cone2}}
\end{figure}


\begin{figure}
\centering{\includegraphics[width=8.5cm,angle=0,clip=]{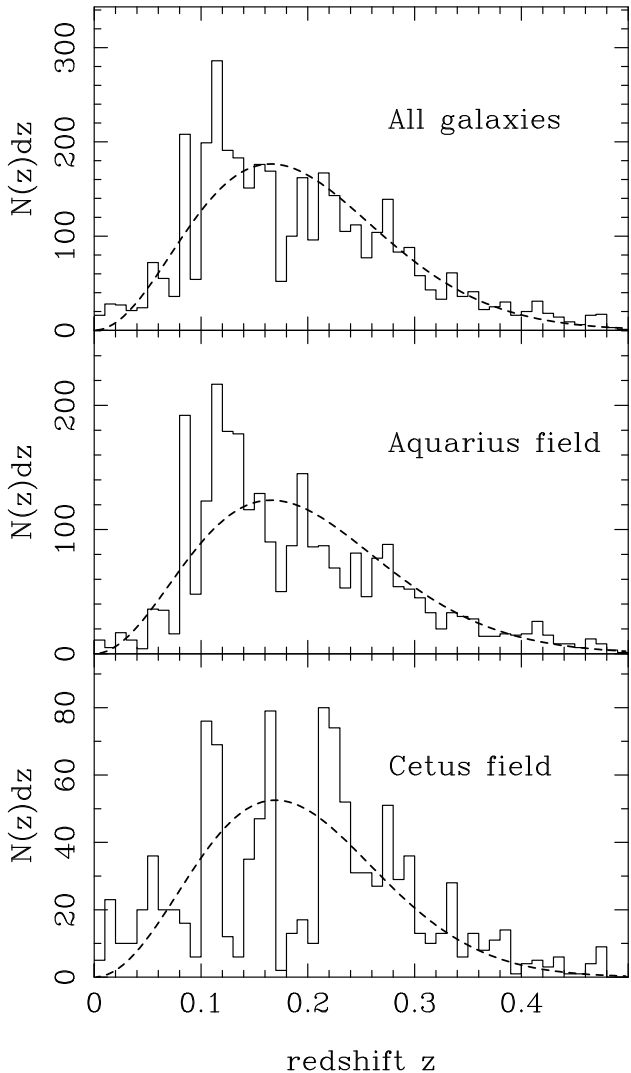}}
\caption{\small{Redshift distributions of the galaxies with
    $Q\ge3$. The top panel represents the total redshift distribution
    for all of the galaxies in the two regions; the middle panel, the
    galaxy distribution in Aquarius; and the bottom panel the galaxy
    distribution in Cetus. The dash line in each panel corresponds to
    an analytic fit to each distribution.}
\label{fig:Nz}}
\end{figure}


\subsection{Spatial Distribution}

Figs~\ref{fig:Cone1} and \ref{fig:Cone2} show the cone plots for all
galaxies with $Q\ge3$ in the Aquarius and Cetus field regions,
respectively. In order to show the large-scale structure more clearly,
the right ascension direction of the plots has been stretched out by a
factor of 10 and thus the apparent transverse filamentary features
should be viewed circumspectly. Nevertheless, it is clear that there
is strong clustering in the distribution of galaxies.  In particular,
the Aquarius region shows strong features at $z\sim$0.08, 0.11, 0.125,
0.14, 0.19 and 0.3, and the Cetus region shows features at
$z\sim0.11$, with also evidence for a void between $0.18<z<0.20$.

Fig.~\ref{fig:Nz} shows the redshift distributions of the galaxies.
Considering the Aquarius field, the $z=0.11$ supercluster established
by \citet{Batuskietal1999} is seen as a spike superposed on a broad
hump between $0.1<z<0.15$. However, there is also another strong spike
at $z\sim0.08$, and as we show in \S7 this represents a further super
cluster.  Considering the Cetus region, there is evidence for three
major structures at $z\sim0.11$, 0.16 and 0.23. Furthermore, the lack
of galaxies between $0.18<z<0.20$ supports the case for the void
indicated in the cone plots.

We have also fitted the commonly used analytic approximation of
\citet{EfstathiouMoody2001} to these distributions, which has the form
\be dN(z)\propto \;z^2
\exp\left[-\left(Bz/\bar{z}\right)^C\right]dz\ , \label{eq:dndz}\ee
where $B$, $C$ are to be determined from the fitting, and $\bar{z}$ is
the median redshift for the distribution. However, from
Fig.~\ref{fig:Nz}, it is clear that this model fails to describe the
complicated features in our distributions. We remark that any lensing
analysis of small field surveys that assumes smooth redshift
distributions as is exemplified by equation (\ref{eq:dndz}), must be
viewed cautiously. For any future analysis where this distribution is
required we will adopt a polynomial fit to the data.


\subsection{Survey Completeness}

\begin{figure*}
  \centering{\includegraphics[width=10.cm,angle=-90,clip=]{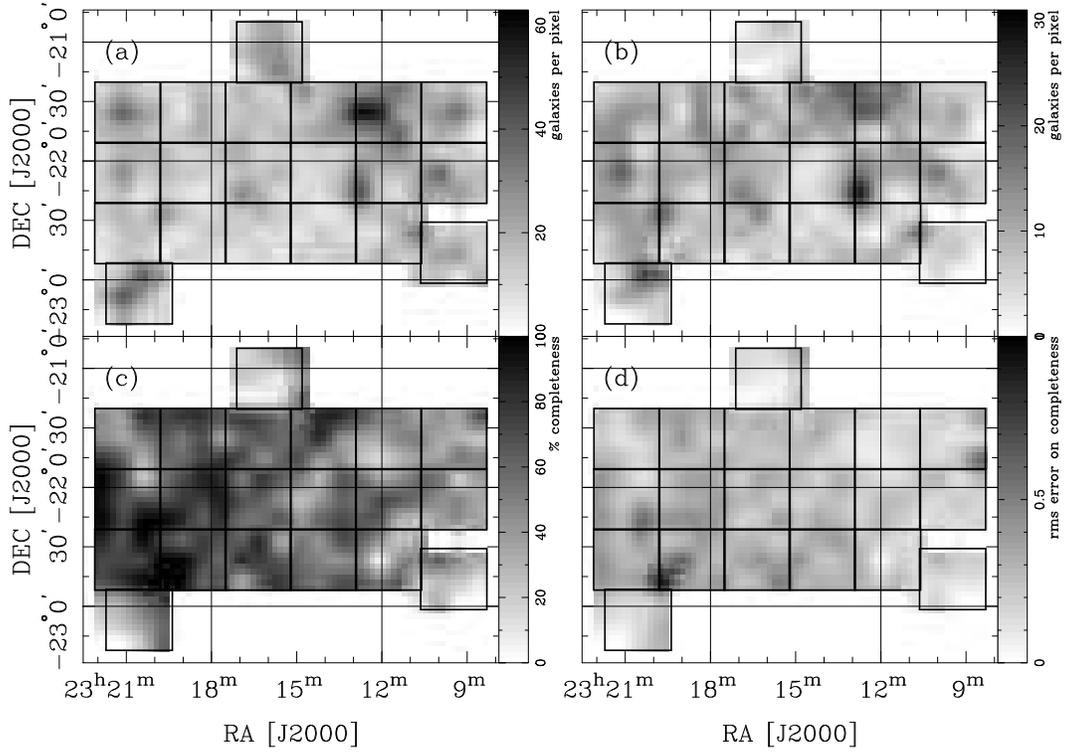}}
  \caption{\small{Bright completeness maps. Panel (a) shows the total
      surface density of galaxies in the target area with R$\le 18.5$
      counted in 10.9$\arcmin$ pixels. Panel (b) shows the surface
      density of bright galaxies with redshift measurements $Q\ge 3$.
      Panel (c) shows the redshift completeness.  Panel (d) shows the
      {\em rms} uncertainty on the completeness estimates.}
    \label{fig:BrightComplete}}
\end{figure*}

\begin{figure*}
  \centering{\includegraphics[width=10cm,angle=-90,clip=]{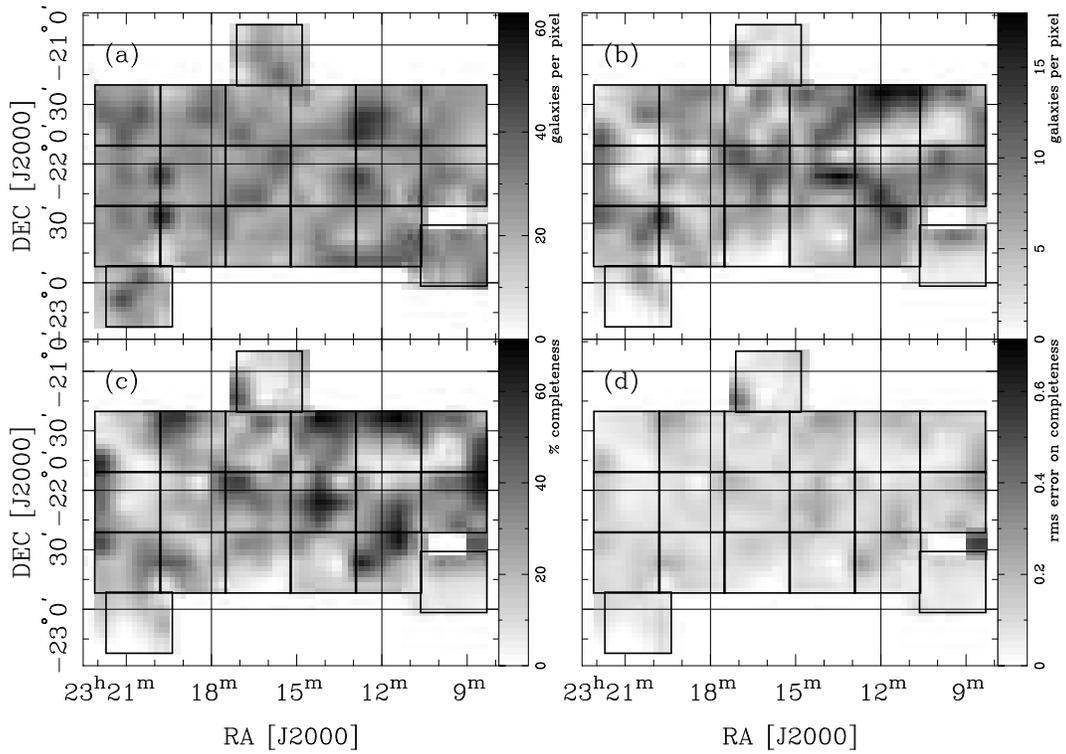}}
  \caption{\small{Identical to Fig.~\ref{fig:BrightComplete}, 
      but for the galaxies with $18.5<$R$\le 19.5$.}
    \label{fig:FaintComplete}}
\end{figure*}

\begin{figure*}
  \centering{\includegraphics[width=10cm,angle=-90,clip=]{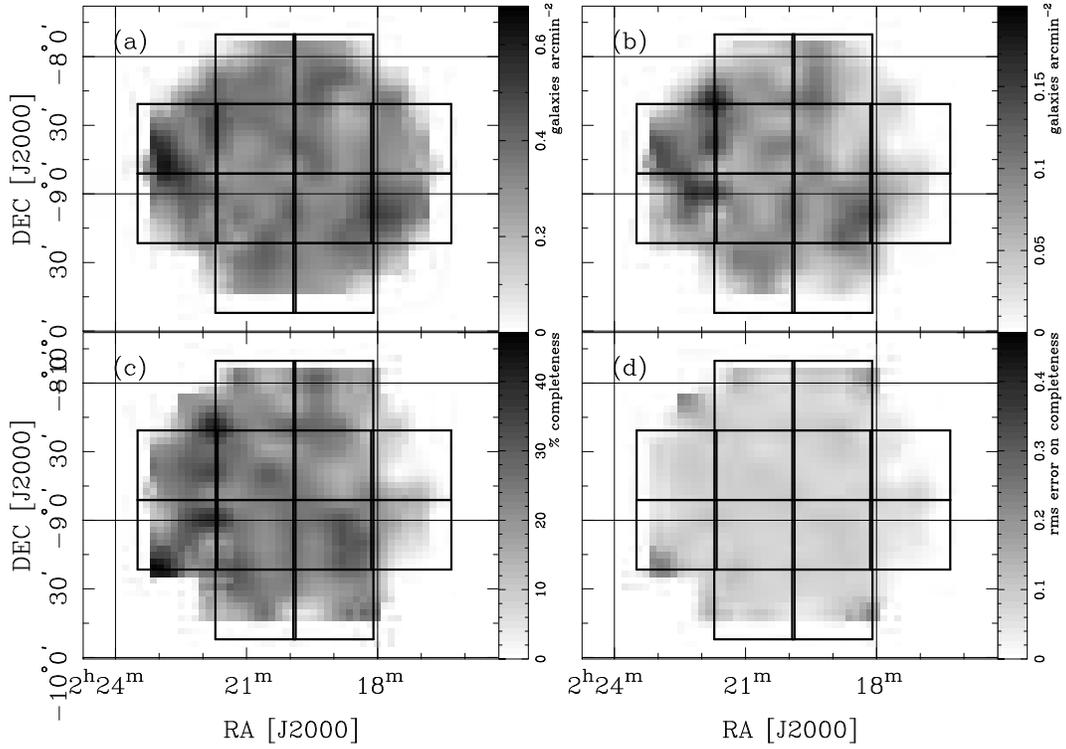}}
  \caption{\small{Cetus region completeness masks 
      for the bright galaxy catalogue: $R>18.5$. Individual panels are
      identical to those of Fig.~\ref{fig:BrightComplete}.}
    \label{fig:ControlBrightComplete}}
\end{figure*}

\begin{figure*}
  \centering{\includegraphics[width=10cm,angle=-90,clip=]{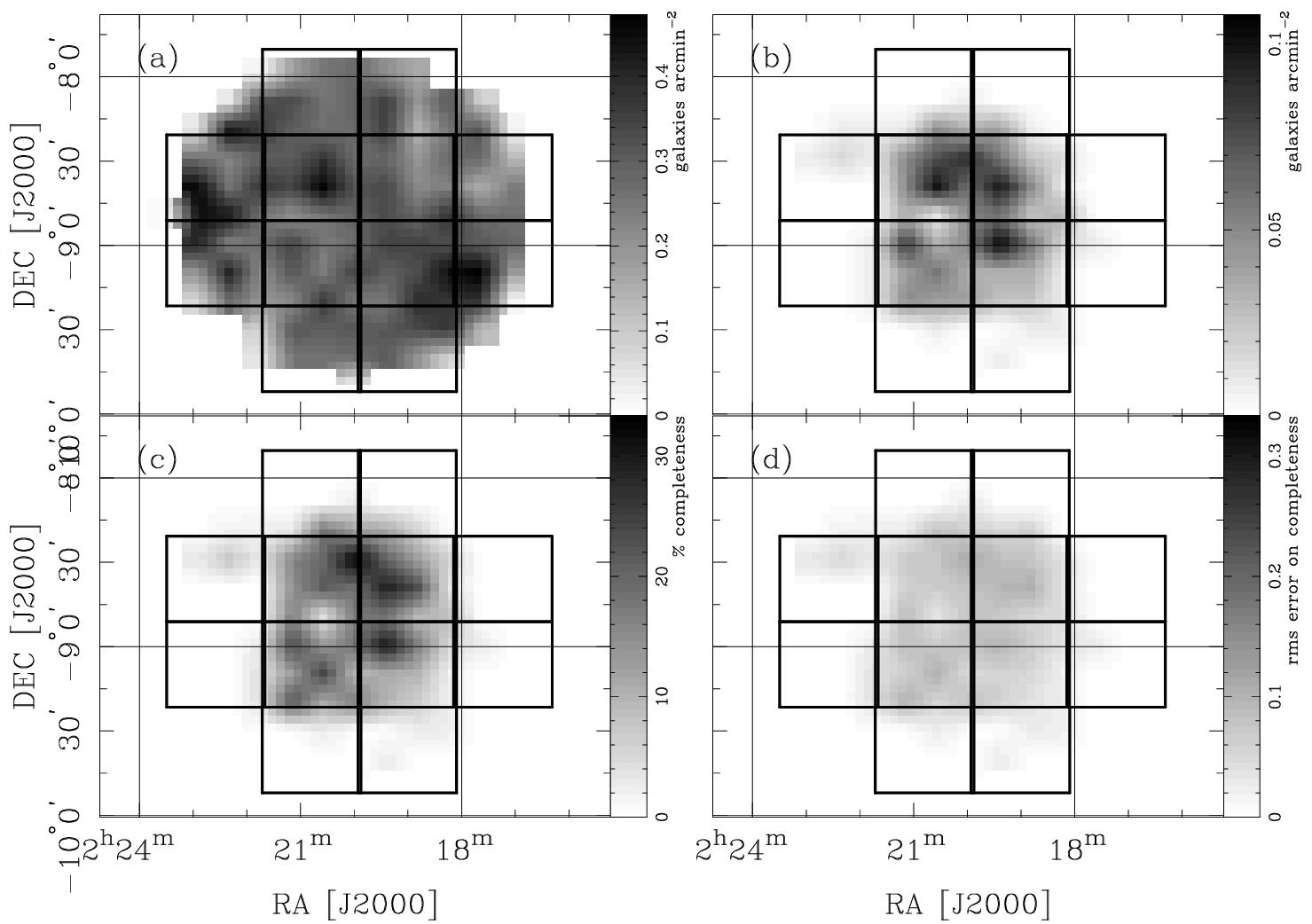}}
  \caption{\small{Cetus field region completeness masks for the faint
      galaxy catalogue: $18.5<$R$\le 19.5$. Individual panels are
      identical to those of Fig.~\ref{fig:BrightComplete}.}
    \label{fig:ControlFaintComplete}}
\end{figure*}

As a final diagnostic, we map out the completeness distributions as a
function of position on the sky. Furthermore, owing to the splitting
of the source catalog into two broad magnitude bins, we generate
completeness masks for the bright and the faint samples separately.
These masks were created as follows: For each magnitude range, the
number of target galaxies and the the number of successful redshifts
were counted in square cells of roughly $~11\arcmin$ on a side, the
completeness for that cell was then estimated as the ratio of these
numbers. The uncertainty on each estimate was determined by combining
the Poisson errors on the two counts in quadrature.  Finally, a local
completeness value was associated to each galaxy through a bi-linear
interpolation of the completeness map corresponding to the galaxy
magnitude. Figs \ref{fig:BrightComplete} and \ref{fig:FaintComplete}
show the completeness distribution for the Aquarius fields, and Figs
\ref{fig:ControlBrightComplete} and \ref{fig:ControlFaintComplete}
show the completeness distribution for the Cetus region.


\section{Cluster Detection}\label{sec:cluster}

We now identify clusters and large groups in the Aquarius and Cetus
regions from the redshift data.  In addition to our 2723 ($Q\ge3$) 2dF
redshifts for Aquarius, we also used 31 redshifts taken from the NED
archives for this analysis. Possible duplicates were rejected by
requiring that any galaxy with a NED redshift should be separated by
$> 5\arcsec$ from any galaxy in our 2dF redshift catalog.  These
literature redshifts were included in the survey completeness
estimates described above.

The cluster candidates were identified as follows: First, we generated
smoothed images of the galaxy distribution. This was done for a series
of redshift planes running from $z_{\rm p} = 0$ to $z_{\rm p} = 0.42$,
separated by $\Delta (cz) = 1000\, {\rm km\, s}^{-1}$.  In generating
these maps each galaxy was weighted by the inverse of the local survey
completeness and Gaussian weighted (with $\sigma = 1000\, {\rm km\,
  s}^{-1}$) according to its difference in radial velocity from the
redshift plane. These frames were all of size $832 \times 832$ pixels
with pixel scale $14.875\arcsec /{\rm pixel}$.  These were then
smoothed in the angular direction with a 2D Gaussian of scale
$33\arcsec / z_{\rm p}$. For the nearest cluster in our catalogue,
$z\sim0.06$, this corresponded to a physical smoothing scale of
$\sim450$\,kpc\,$h^{-1}$ and for the most distant cluster at
$z\sim0.3$, a scale of $\sim350$\,kpc\,$h^{-1}$.


Next, we located the peaks in the smoothed frames. This was done using
the peak finder algorithm in the IMCAT software package
\citep{Kaiseretal1995}. In accord with the image generation, we
detected peaks using a Gaussian smoothing filter with a radius of $33
\arcsec /z_{\rm p}$. The peak detection threshold was set to be a low
significance threshold $\nu = 2$. A ``galaxy flux'' was then
associated to each of the identified peaks, by summing over all pixels
within $33\arcsec /z_{\rm p}$ from the central pixel. All peaks below
a flux threshold of 375 galaxies were rejected.  This cutoff level was
arrived at by determining the highest peak height that would allow us
to recover all of the previously known clusters in the fields. In
practice, our detection scheme would typically recover the same
structure in several adjacent redshift planes. We rejected such
duplicates, by identifying those peaks that had a corresponding peak
with higher flux in a neighbouring plane and that were separated by
less than one Abell radius $R_A = 1\arcmin 7 / z$.

As a first approximation, the redshift of each candidate cluster was
assumed to be that of the plane $z_{\rm p}$.  We then defined as
cluster members all galaxies within a radial velocity interval of
width $\Delta (cz) = 6000\, {\rm km\, s}^{-1}$, centered on the
estimated cluster redshift.  Finally, we generated a cluster catalog
containing clusters that had $N\ge9$ members with spectroscopically
measured redshifts.  This selection criterion was motivated by the
desire to have enough redshifts to make at least a rough estimate of
the cluster velocity dispersion (see the following section). The final
cluster catalogs contain a total of 48 identified objects in Aquarius
(see Table~\ref{tab:AQRlocation}) and 14 clusters in Cetus (see
Table~\ref{tab:CETlocation}).

In these Tables, column (1) is the identifier in our catalog, where we
have adopted the naming conventions ``AQ HHMM.M(+/-)DDMM'' for the
Aquarius clusters and ``CET HHMM.M(+/-)DDMM'' for the Cetus clusters.
The last nine digits are the celestial coordinates of the objects,
which we take to be the position of the peaks in the smoothed galaxy
distributions.  Here, the right ascension is given in units of hours
and minutes, with one decimal, and the declination is given in units
of degrees and arc-minutes. The precision of these coordinates may
vary significantly for different clusters because of variations in
survey completeness as a function of sky position (see Figs
\ref{fig:BrightComplete} -- \ref{fig:ControlFaintComplete}).
Furthermore, we note that the minimum distance between the 2dF fibres
limits our spectroscopic sampling in the densest parts of the
clusters, and this limits the precision of positional estimates based
on our spectroscopic data only. Such sampling restrictions will also
limit the detection efficiency for the more distant clusters, which
cover a smaller sky area.  An updated catalog with more accurate
coordinates for the center of each cluster will be reported in a
future paper, based on our $B$- and $R$-band imaging data from the ESO
WFI.


\subsection{Aquarius Region}

In column (2) of Table~\ref{tab:AQRlocation} we give the ACO cluster
catalog name or if the cluster does not have an ACO identifier, then
we give the AqrCC catalog name (Caretta02).  Of the 23 previously
known clusters in our Aquarius survey region, we are able to robustly
identify 20, of which five had previously unknown redshifts and two
appear to be multiple systems.  In addition to these, two previously
known clusters (\objectname{A2541} and \objectname{A2550}) are
recovered by our detection method when we include 58 literature
redshifts for galaxies that are just outside our survey region and
that are fainter than $R=19.5$ (see next section for details).  The
only previously known system that we fail to detect is the cluster
candidate \objectname{AqrCC~43}, which is a peak in the 2D galaxy
distribution on the sky (Caretta02), located close to the Northern
edge of our survey region.

\begin{figure*}
\centering{\includegraphics[width=12cm,angle=-90,clip=]{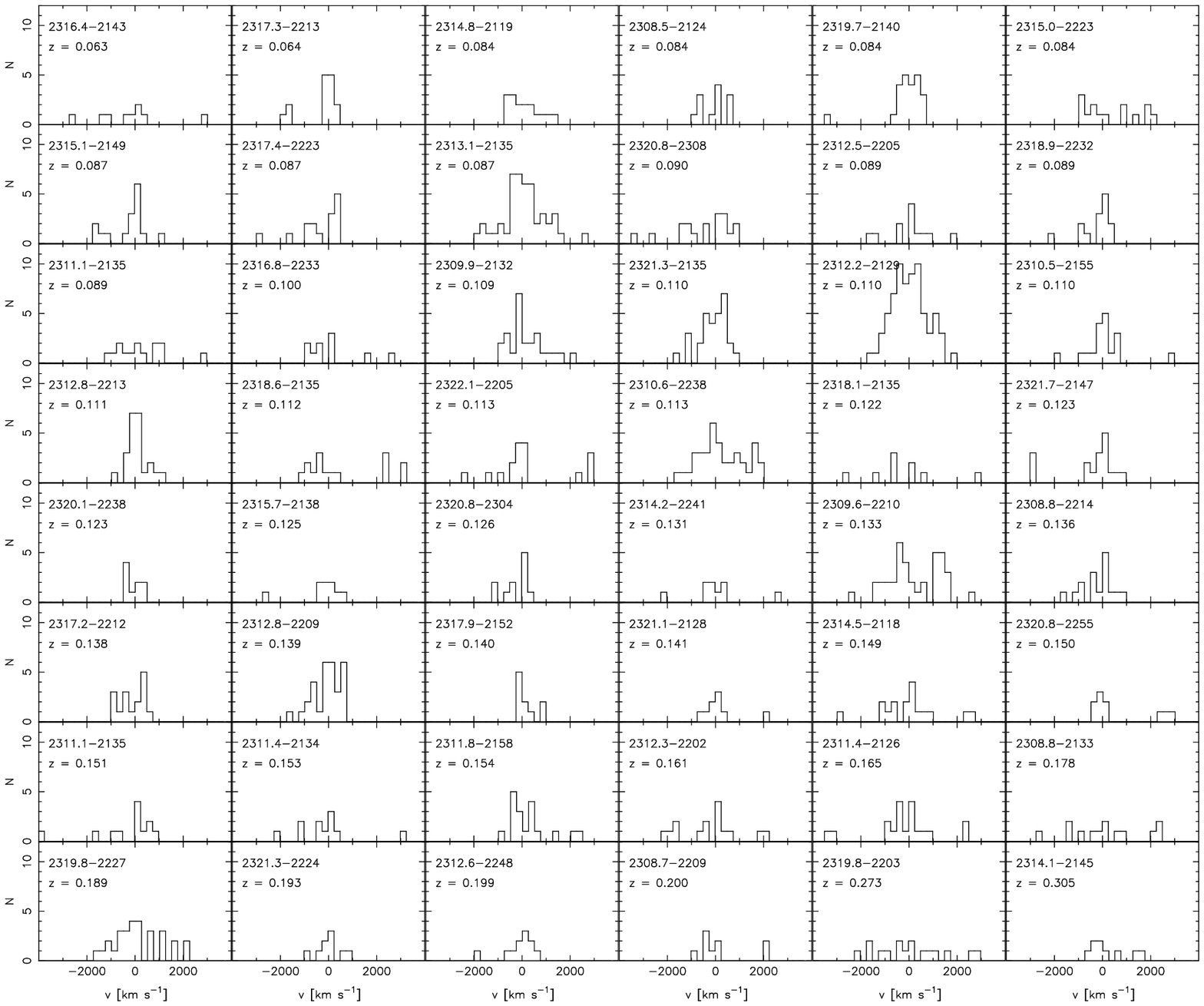}}
\caption{\small{Velocity histograms for the 48 clusters identified in
the Aquarius region. The properties of each cluster are listed in
Table \ref{tab:AQRlocation}}\label{fig:velhist_AQR}}
\end{figure*}

\begin{figure*}
\centering{\includegraphics[width=7cm,angle=-90,clip=]{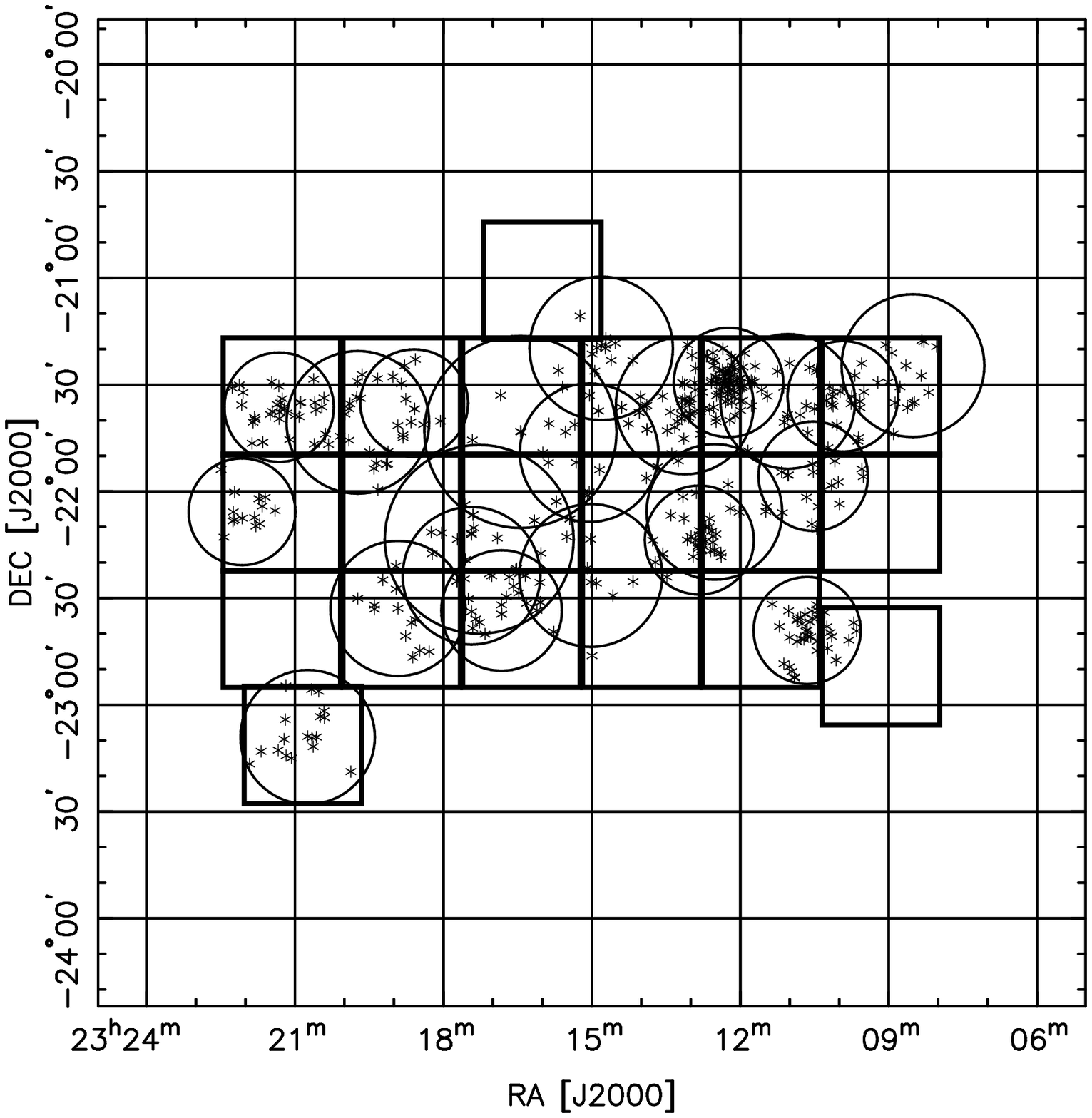}
\includegraphics[width=7cm,angle=-90,clip=]{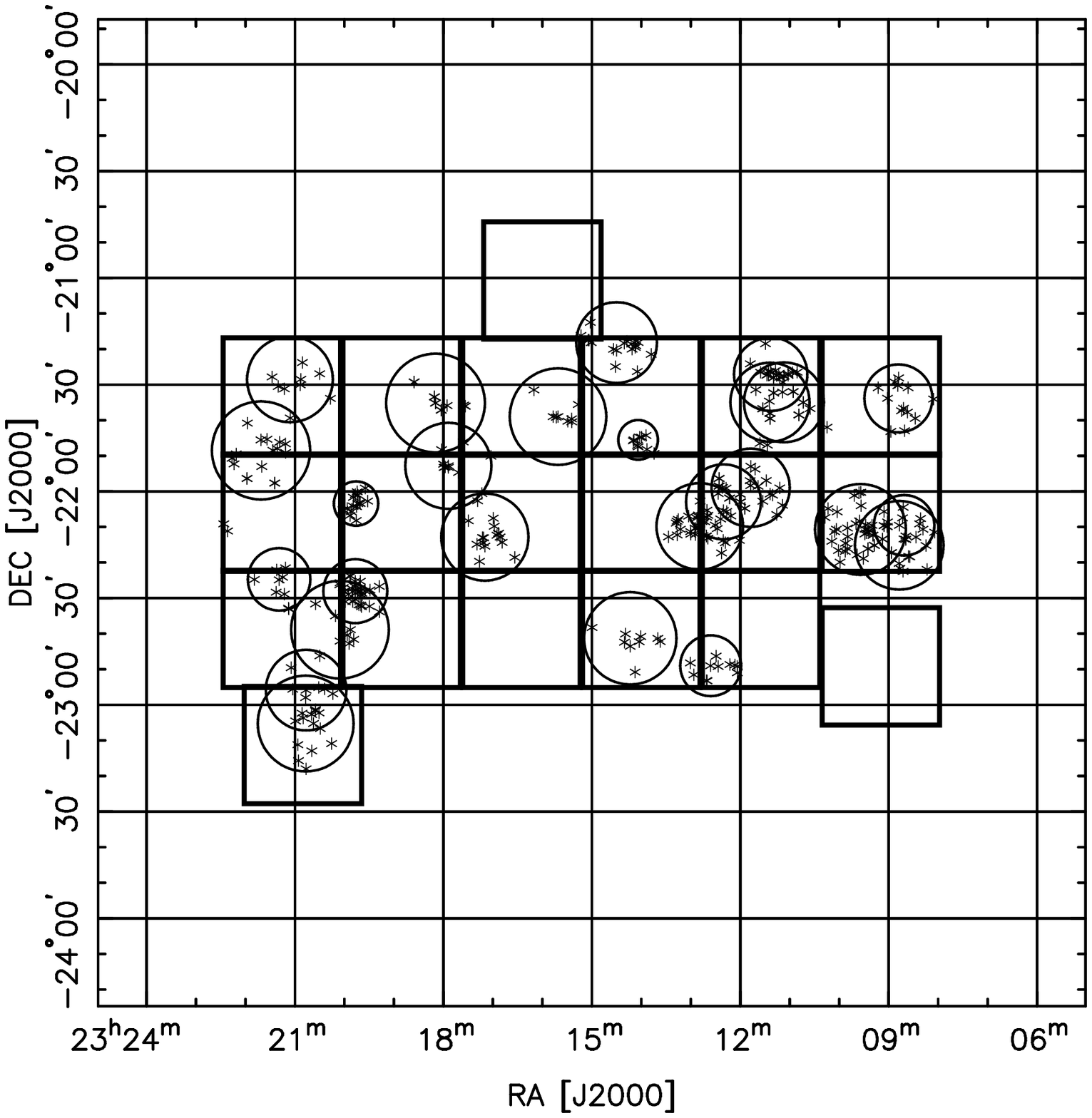}}
\caption{\small{2D plot of the distribution of the identified clusters
in the Aquarius region (left panel: Clusters with $z < 0.12$; right
panel: Clusters with $z > 0.12$).  The member galaxies for all
clusters with $N \ge 9$ spectroscopically determined galaxy redshifts
are indicated by the asterisks, and the Abell radius $R_A$ of each
cluster is indicated by the circles.}
\label{fig:2Dclusters}}
\end{figure*}

Table~\ref{tab:AQRlocation} also contains data for 26 additional
clusters that are previously unknown
systems. Figure~\ref{fig:velhist_AQR} shows radial velocity histograms
for the 48 clusters identified with the methods described here.
Figure~\ref{fig:2Dclusters} shows the cluster distribution in Aquarius
projected on the sky. Each cluster is represented by a circle, where
the scale of the circle is proportional to the Abell radius for the
cluster.


\begin{deluxetable*}{clccccccc}
\tabletypesize{\small}
\tablecolumns{6}
  \tablewidth{0pt}
  \tablecaption{Cluster Redshifts and Velocity
  Dispersions for the Aquarius Region\label{tab:AQRlocation}}   
  \tablehead{
  \colhead{ID} & \colhead{other ID} & \colhead{$N$} &   \colhead{$z$} & 
  \colhead{$N_{\rm lit}$\tablenotemark{a}} &\colhead{$z_{\rm lit}$\tablenotemark{a}} & 
  \colhead{$\sigma_P$ (${\rm km}\,{\rm s}^{-1} $)} & \colhead{Asym.\ index} & 
  \colhead{Tail index} \\
  \colhead{(1)} & \colhead{(2)} & \colhead{(3)} & \colhead{(4)} & 
  \colhead{(5)} &\colhead{(6)} & 
  \colhead{(7)} & \colhead{(8)} &
  \colhead{(9)} 
   }\startdata
   \objectname{AQ2308.5-2124} & \nodata
  & 12 & $0.0847^{+0.0005}_{-0.0006}$ & \nodata & \nodata &
  $520^{+90}_{-46}$ & -0.212 & 0.705 \\
   \objectname{AQ2308.7-2209} &
  \nodata & 9 & $0.1993^{+0.0003}_{-0.0005}$ & \nodata & \nodata &
  $382^{+816}_{-160}$ & -0.007 & 1.281 \\
   \objectname{AQ2308.8-2133} &
   \objectname{A2539} & 13 & $0.1782^{+0.0014}_{-0.0011}$ & 4 & 0.1863 &
  $1383^{+317}_{-117}$ & -0.522 & 0.980 \\
   \objectname{AQ2308.8-2214} &
  \nodata & 16 & $0.1361^{+0.0005}_{-0.0007}$ & \nodata & \nodata &
  $614^{+150}_{-81}$ & -0.207 & 1.054 \\
   \objectname{AQ2309.6-2210}
  & \objectname{A2540} & 38 & $0.1340^{+0.0006}_{-0.0006}$ & 8 & 0.129
  & $978^{+120}_{-77}$ & -0.643 & 0.800 \\
  \objectname{AQ2309.9-2132} & \objectname{AqrCC 24} & 25 &
  $0.1097^{+0.0006}_{-0.0004}$ & 7 & 0.1109 &
 $701^{+156}_{-84}$ &
  0.902 & 1.114 \\ 
   \objectname{AQ2310.5-2155} &
  \nodata & 18 & $0.1107^{+0.0003}_{-0.0002}$ & \nodata & \nodata &
  $389^{+298}_{-101}$ & 0.024 & 1.522 \\
  \objectname{AQ2310.6-2238} & \objectname{A2546} &
  38 & $0.1136^{+0.0006}_{-0.0005}$ & 22 & 0.113 & $901^{+88}_{-70}$
  & -0.992 & 0.955 \\
   \objectname{AQ2311.1-2134} &
  \nodata & 12 & $0.1515^{+0.0006}_{-0.0012}$ & \nodata & \nodata &
  $770^{+795}_{-180}$ & -1.102 & 1.339 \\
  \objectname{AQ2311.1-2135} &
  \nodata & 14 & $0.0903^{+0.0010}_{-0.0009}$ & \nodata & \nodata &
  $957^{+388}_{-136}$ & 0.856 & 0.777 \\
   \objectname{AQ2311.4-2126} &
  \objectname{AqrCC 32a} & 19 & $0.1647^{+0.0007}_{-0.0004}$ & \nodata
  & \nodata & $912^{+613}_{-293}$ & -0.215 & 2.271 \\
   \objectname{AQ2311.4-2134} &
  \objectname{AqrCC 32b} & 11 & $0.1520^{+0.0006}_{-0.0010}$ & \nodata & \nodata &
  $621^{+647}_{-231}$ & 0.171 & 1.686 \\
   \objectname{AQ2311.8-2158} &
  \nodata & 18 & $0.1545^{+0.0006}_{-0.0005}$ & \nodata & \nodata &
  $562^{+343}_{-165}$ & 0.783 & 1.574 \\
   \objectname{AQ2312.2-2129} &
  \objectname{A2554} & 71 & $0.1106^{+0.0003}_{-0.0003}$ & 35 & 0.1108 &
  $680^{+68}_{-43}$ & 0.297 & 1.074 \\
   \objectname{AQ2312.3-2202} &
  \nodata & 15 & $0.1601^{+0.0007}_{-0.0012}$ & \nodata & \nodata &
  $1123^{+238}_{-176}$ & 0.319 & 1.125 \\
   \objectname{AQ2312.5-2205} &
  \nodata & 12 & $0.0891^{+0.0004}_{-0.0010}$ & \nodata & \nodata &
  $820^{+325}_{-216}$ & 0.431 & 1.627 \\
   \objectname{AQ2312.6-2248} & \objectname{AqrCC 34}
  & 11 & $0.1990^{+0.0004}_{-0.0004}$ & \nodata & \nodata &
  $374^{+301}_{-83}$ & -0.758 & 1.408 \\
   \objectname{AQ2312.8-2209} &
  \nodata & 30 & $0.1397^{+0.0003}_{-0.0003}$ & \nodata & \nodata &
  $507^{+91}_{-50}$ & -1.088 & 0.956 \\
   \objectname{AQ2312.8-2213} &
  \objectname{A2555} & 23 & $0.1109^{+0.0002}_{-0.0002}$ & 11 & 0.1106
  & $366^{+120}_{-99}$ & 0.172 & 1.628 \\
   \objectname{AQ2313.1-2135} &
  \objectname{A2556} & 46 & $0.0881^{+0.0004}_{-0.0004}$ & 9 & 0.0871 &
  $816^{+137}_{-99}$ & 0.068 & 1.434 \\
   \objectname{AQ2314.1-2145} &
  \nodata & 9 & $0.3047^{+0.0009}_{-0.0002}$ & \nodata & \nodata &
  $447^{+314}_{-333}$ & 0.736 & 1.065 \\
   \objectname{AQ2314.2-2241}
  & \nodata & 9 & $0.1306^{+0.0005}_{-0.0004}$ & \nodata & \nodata &
  $288^{+1111}_{-71}$ & -0.198 & 3.138 \\
   \objectname{AQ2314.5-2118} &
  \nodata & 17 & $0.1495^{+0.0007}_{-0.0007}$ & \nodata & \nodata &
  $1048^{+463}_{-239}$ & -0.081 & 1.412 \\
   \objectname{AQ2314.8-2119}
  & \objectname{A2565} & 16 & $0.0845^{+0.0006}_{-0.0005}$ & 12 &
  0.0825 & $584^{+114}_{-63}$ & 1.155 & 1.130 \\
   \objectname{AQ2315.0-2223} &
  \nodata & 14 & $0.0857^{+0.0012}_{-0.0017}$ & \nodata & \nodata &
  $1077^{+228}_{-74}$ & 1.160 & 0.728 \\
   \objectname{AQ2315.1-2149} &
  \nodata & 16 & $0.0873^{+0.0003}_{-0.0003}$ & \nodata & \nodata &
  $478^{+313}_{-306}$ & -0.180 & 1.083 \\
  \objectname{AQ2315.7-2138} & \nodata & 9 &
  $0.1253^{+0.0004}_{-0.0003}$ & \nodata & \nodata &
  $309^{+381}_{-67}$ & -0.697 & 2.161 \\
   \objectname{AQ2316.4-2143} &
  \nodata & 9 & $0.0620^{+0.0008}_{-0.0025}$ & \nodata & \nodata &
  $1016^{+1017}_{-553}$ & 0.039 & 1.655 \\
   \objectname{AQ2316.8-2233} &
  \nodata & 10 & $0.0995^{+0.0007}_{-0.0010}$ & \nodata & \nodata &
  $826^{+630}_{-323}$ & 1.107 & 1.577 \\
   \objectname{AQ2317.2-2212} &
  \objectname{A2568} & 16 & $0.1383^{+0.0005}_{-0.0005}$ & 6 & 0.1397
  & $488^{+75}_{-41}$ & -0.443 & 0.724 \\
   \objectname{AQ2317.3-2213} &
  \nodata & 15 & $0.0643^{+0.0002}_{-0.0002}$ & \nodata & \nodata &
  $152^{+187}_{-24}$ & -1.126 & 3.054 \\
   \objectname{AQ2317.4-2223} &
  \objectname{AqrCC 48a} & 15 & $0.0874^{+0.0009}_{-0.0008}$ & 5 & 0.0827 &
  $654^{+401}_{-143}$ & -1.484 & 1.178 \\
   \objectname{AQ2317.9-2152} &
  \nodata & 10 & $0.1405^{+0.0006}_{-0.0002}$ & \nodata & \nodata &
  $314^{+182}_{-139}$ & 1.341 & 1.222 \\
   \objectname{AQ2318.1-2135} &
  \nodata & 9 & $0.1209^{+0.0013}_{-0.0009}$ & \nodata & \nodata &
  $1202^{+743}_{-343}$ & 0.257 & 2.025 \\
   \objectname{AQ2318.6-2135} &
  \nodata & 15 & $0.1115^{+0.0032}_{-0.0009}$ & \nodata & \nodata &
  $1586^{+511}_{-261}$ & 0.070 & 0.678 \\
   \objectname{AQ2318.9-2232} &
  \objectname{AqrCC 48b} & 14 & $0.0894^{+0.0003}_{-0.0007}$ & 5 & 0.0827 &
  $417^{+261}_{-126}$ & -0.851 & 0.940 \\
   \objectname{AQ2319.7-2140} &
  \objectname{AqrCC 56} & 23 & $0.0849^{+0.0003}_{-0.0003}$ & \nodata & \nodata &
  $340^{+46}_{-32}$ & -0.490 & 0.755 \\
   \objectname{AQ2319.8-2203}
  & \objectname{A2575} & 16 & $0.2731^{+0.0011}_{-0.0010}$ & \nodata &
  \nodata & $1200^{+243}_{-155}$ & 0.016 & 1.134 \\
   \objectname{AQ2319.8-2227} &
  \objectname{A2576} & 29 & $0.1895^{+0.0006}_{-0.0006}$ & 9 & 0.1876 &
  $887^{+131}_{-87}$ & 0.048 & 1.047 \\
   \objectname{AQ2320.1-2238} &
  \nodata & 9 & $0.1235^{+0.0004}_{-0.0005}$ & \nodata & \nodata &
  $305^{+115}_{-25}$ & 0.440 & 0.854 \\
   \objectname{AQ2320.8-2255} &
  \nodata & 10 & $0.1496^{+0.0003}_{-0.0002}$ & \nodata & \nodata &
  $182^{+1162}_{-41}$ & 1.048 & 0.610 \\
   \objectname{AQ2320.8-2304} &
  \objectname{A2577} & 11 & $0.1256^{+0.0006}_{-0.0009}$ & 7 & 0.1248 &
  $496^{+252}_{-59}$ & -0.584 & 1.222 \\
  \objectname{AQ2320.8-2308} & \objectname{A2580} & 18 &
  $0.0890^{+0.0009}_{-0.0010}$ & 17 & 0.0890 & $1013^{+370}_{-151}$ &
  -0.909 & 1.058 \\
   \objectname{AQ2321.1-2128} &
  \nodata & 9 & $0.1410^{+0.0002}_{-0.0004}$ & \nodata & \nodata &
  $264^{+554}_{-73}$ & 0.599 & 3.069 \\
   \objectname{AQ2321.3-2135} &
  \objectname{A2579} & 30 & $0.1103^{+0.0004}_{-0.0004}$ & 9 & 0.1114 &
  $543^{+98}_{-61}$ & -0.706 & 1.110 \\
   \objectname{AQ2321.3-2224} & \objectname{AqrCC
  62} & 9 & $0.1935^{+0.0004}_{-0.0005}$ & \nodata & \nodata &
  $417^{+138}_{-80}$ & 0.010 & 1.359 \\
   \objectname{AQ2321.7-2147} &
  \nodata & 17 & $0.1233^{+0.0003}_{-0.0003}$ & \nodata & \nodata &
  $416^{+648}_{-129}$ & -0.085 & 2.382 \\
   \objectname{AQ2322.1-2205} &
  \objectname{A3996} & 17 & $0.1125^{+0.0003}_{-0.0011}$ & 6 & 0.0854 &
  $841^{+918}_{-486}$ & -0.625 & 1.634 \\  
 \enddata

\tablenotetext{a}{The literature redshifts are all taken from Caretta02.}
\tablecomments{See the text for further comments on specific clusters.}

\end{deluxetable*}


\subsection{Cetus Region}

\begin{figure}
\centering{\includegraphics[width=5.6cm,angle=-90,clip=]{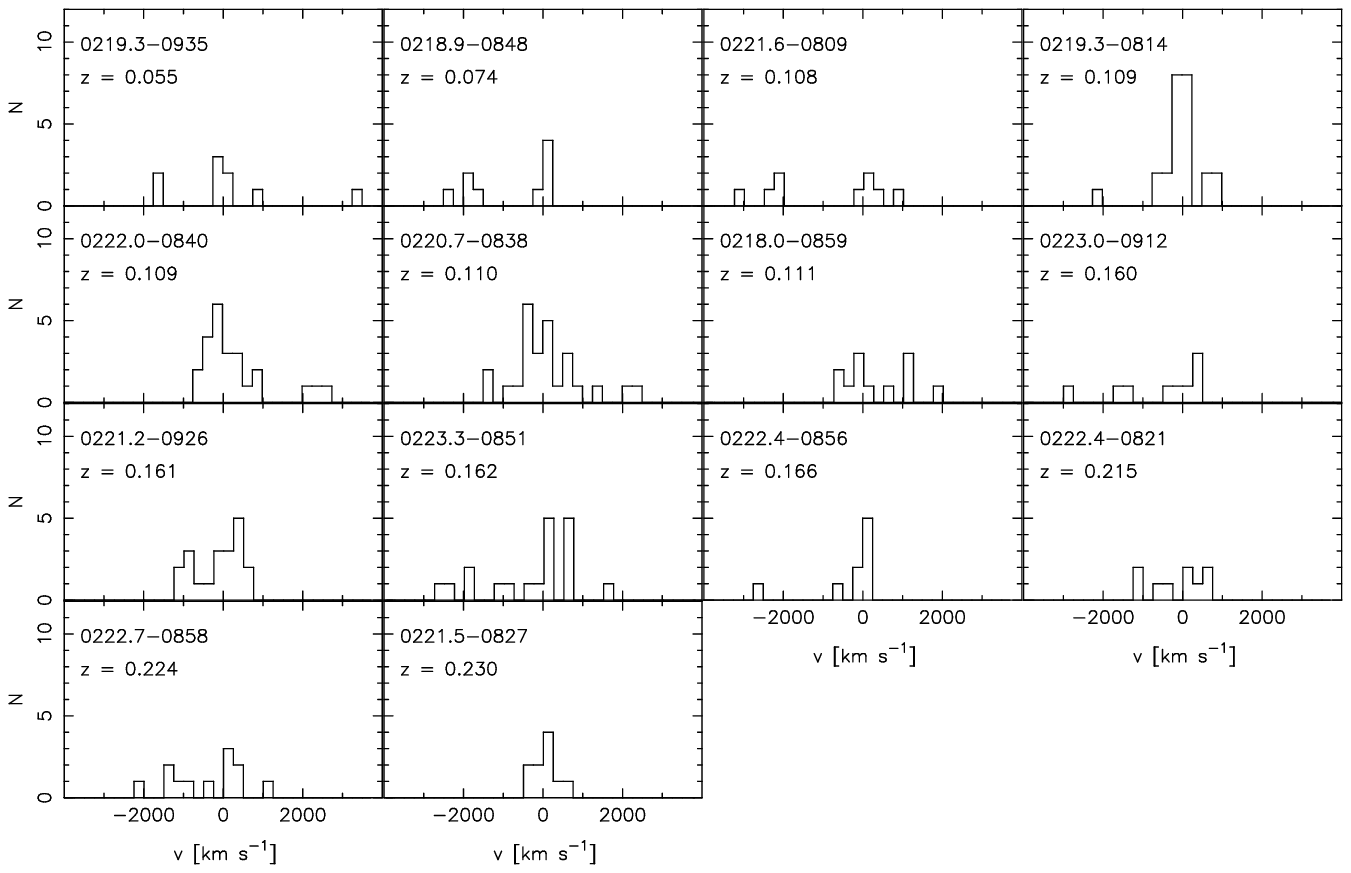}}
\caption{\small{Velocity histograms for the 14 clusters identified in
the Cetus region. The properties of each cluster are listed in Table
\ref{tab:CETlocation}}\label{fig:velhist_CET}}
\end{figure}

\begin{figure}
\centering{\includegraphics[width=7cm,angle=-90,clip=]{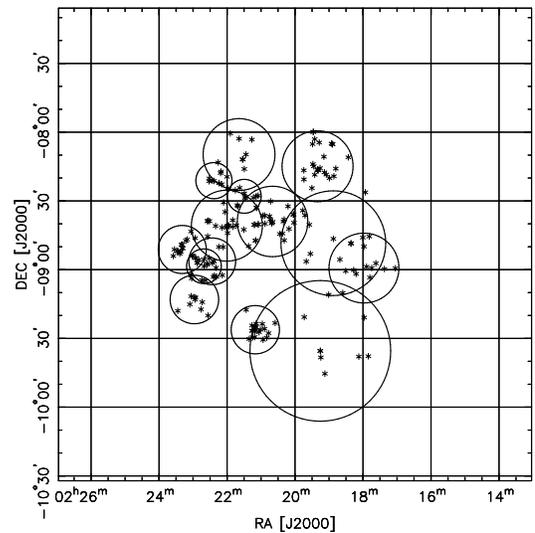}}
\caption{\small{2D plot of the distribution of the identified clusters
in the Cetus region.  The member galaxies for all clusters with $N \ge
9$ spectroscopically determined galaxy redshifts are indicated by the
asterisks, and the Abell radius $R_A$ of each cluster is indicated by
the circles.}
\label{fig:2Dclusters_CET}}
\end{figure}

In the typical field region in Cetus, our cluster detection method
recovers 12 new galaxy clusters in addition to the 2 previously known
clusters identified in column (2) of Table~\ref{tab:CETlocation}.
Figure~\ref{fig:velhist_CET} shows radial velocity histograms for all
14 clusters.  Figure~\ref{fig:2Dclusters_CET} shows the cluster
distribution projected on the sky. Each cluster is represented by a
circle, where the scale of the circle is proportional to the Abell
radius for the cluster.


\section{Cluster Redshifts and Velocity Dispersions}\label{sec:velocity}

We now present estimates of the centroid redshifts and the velocity
dispersions for the cluster candidates. Owing to the modest number of
observed galaxies per cluster, typically between 9 and 71, it was
important to use robust statistical estimators to avoid problems
caused by e.g., deviations from a Gaussian velocity distribution and
contamination from fore- and background galaxies. We therefore use the
biweight estimators of location and scale described by
\citet{Beersetal1990}, to measure velocity dispersions and centroid
redshifts.
%
%
Confidence intervals for the estimates were calculated using the bias
corrected and accelerated bootstrap method \citep{Efron1987} with
10,000 bootstrap replications. These robust estimators and methods for
determining confidence intervals have been well tested on redshift
samples like ours \citep[see
e.g.,][]{Beersetal1990,Girardietal1993,Borganietal1999,Irgensetal2002}.
Our statistical analysis was performed using the {\it ROSTAT} program
kindly provided by Dr.\ T.\ C.\ Beers.

The cluster redshifts and one-dimensional velocity dispersion
estimates, together with 68\% confidence limits, are presented for the
Aquarius region in Table~\ref{tab:AQRlocation} as columns (4) and (7)
and for the Cetus region in Table~\ref{tab:CETlocation} as columns (4)
and (6). Figures \ref{fig:Cone1} and \ref{fig:Cone2} show the cluster
candidates overlayed on the measured galaxy distribution.

Column (3) in both tables shows the number of galaxy redshifts used in
these estimates for each cluster. We note that from the sample, there
are 15 clusters for which estimates were derived from only 9
redshifts, and as expected these have respectively large confidence
limits for the velocity dispersion estimates.  Previous redshift
estimates have been made by Caretta02 for a total of 15 clusters in
our Aquarius survey region (one of which we identify as a double
system).  These values are listed in column (6) of
Table~\ref{tab:AQRlocation}, and the number of cluster members used
for their redshift estimate is given in column (5).  We find very good
agreement with the literature redshifts for 12 clusters, while 3
clusters have discrepant redshift values. These cases are discussed in
detail below.  However, we point out that our estimates are based upon
more (in most cases many more) galaxies than Caretta02. In the Cetus
region, \citet{Romeretal2000} have published a redshift for a single
cluster, listed in column (5) of Table~\ref{tab:CETlocation}, which is
consistent with our value.

Columns (8) and (9) of Table~\ref{tab:AQRlocation} and columns (7) and
(8) of Table~\ref{tab:CETlocation} show the asymmetry index and
normalized tail index defined by \citet{BirdBeers1993}. For a Gaussian
distribution, the asymmetry index is zero and the normalized tail
index is unity. Large deviations from these numbers are a strong
indication that the cluster is not in dynamic equilibrium, or that it
is strongly contaminated by non-cluster galaxies, i.e., that the found
velocity dispersion is not to be taken as a good measurement of its
dynamical mass. For our sample sizes, an asymmetry index with absolute
value larger than 0.9 and/or a reduced tail index larger than 1.5
(somewhat higher values for the clusters with $N<25$) should signify
that the cluster velocity dispersion is suspect. We see that the
bootstrap method gives large confidence limits on the velocity
dispersions for these clusters.


\begin{deluxetable*}{clccccccc}

\tablecolumns{6}
  \tablewidth{0pt} 
\tablecaption{Cluster Redshifts and Velocity
  Dispersions for the Cetus Region\label{tab:CETlocation}} 

\tablehead{ \colhead{ID} &
  \colhead{other ID} & \colhead{$N$} & \colhead{$z$} &
  \colhead{$z_{\rm
  lit}$\tablenotemark{a}} & \colhead{$\sigma_P$ (${\rm km}\,{\rm
  s}^{-1} $)} & \colhead{Asym.\ index} & \colhead{Tail index} \\
  \colhead{(1)} & \colhead{(2)} & \colhead{(3)} & \colhead{(4)} &
  \colhead{(5)} &\colhead{(6)} & \colhead{(7)} & \colhead{(8)} }
 
\startdata

\objectname{CET0218.0-0859} & A339 & 12 &
  $0.1123^{+0.0011}_{-0.0012}$ & \nodata & $778^{+179}_{-82}$
  & 0.436 & 0.735 \\ 

\objectname{CET0218.9-0848} & \nodata & 9 &
  $0.0745^{+0.0002}_{-0.0001}$ & \nodata &
  $118^{+1065}_{-41}$ & -0.843 & 0.650 \\ 

\objectname{CET0219.3-0814} &  \nodata & 25 &
$0.1097^{+0.0002}_{-0.0002}$ 
& \nodata & $342^{+112}_{-79}$ & -0.317 & 2.026 \\ 

\objectname{CET0219.3-0935} &
  \nodata & 9 & $0.0551^{+0.0004}_{-0.0006}$ & \nodata &
  $437^{+886}_{-381}$ & 0.710 & 2.313 \\ 

\objectname{CET0220.7-0838} & \nodata & 26 &
$0.1102^{+0.0004}_{-0.0004}$ & \nodata & $697^{+236}_{-146}$
& 0.277 & 1.375 \\

\objectname{CET0221.2-0926} & \nodata & 20 &
  $0.1609^{+0.0005}_{-0.0005}$ & \nodata & $526^{+72}_{-41}$ &
  -0.034 & 0.738 \\ 

\objectname{CET0221.6-0809} & \nodata &
  9 & $0.1056^{+0.0026}_{-0.0040}$ & \nodata &
  $1484^{+542}_{-115}$ & -0.205 & 0.811 \\ 

\objectname{CET0221.5-0827} &
  \nodata & 10 & $0.2302^{+0.0002}_{-0.0003}$ & \nodata &
  $251^{+74}_{-45}$ & 0.179 & 1.307 \\ 

\objectname{CET0222.0-0840} &
  \nodata & 24 & $0.1097^{+0.0005}_{-0.0003}$ & \nodata
  & $589^{+404}_{-133}$ & 1.028 & 1.584 \\

\objectname{CET0222.4-0821} & \nodata & 9 &
  $0.2152^{+0.0006}_{-0.0007}$ & \nodata &
  $569^{+149}_{-39}$ & -1.223 & 0.718 \\ 

\objectname{CET0222.4-0856} & \nodata & 9 &
  $0.1668^{+0.0002}_{-0.0001}$ & \nodata &
  $165^{+283}_{-56}$ & -1.435 & 3.144 \\ 

\objectname{CET0222.7-0858} &
  \nodata & 12 & $0.2234^{+0.0012}_{-0.0011}$ &\nodata
  & $780^{+210}_{-83}$ & -0.487 & 0.489 \\ 

\objectname{CET0223.0-0912}
  & \nodata & 9 & $0.1598^{+0.0008}_{-0.0028}$ & \nodata &
  $918^{+567}_{-266}$ & -1.023 & -0.907 \\ 

\objectname{CET0223.3-0851} &
  RX J0223.4-0852 & 19 & $0.1615^{+0.0007}_{-0.0014}$ & 0.163
  & $1035^{+220}_{-170}$ & -0.705 & 0.874 \\ 

\enddata

\tablenotetext{a}{The literature redshift is taken from
\citet{Romeretal2000}.}
\end{deluxetable*}


We have made some notes on specific clusters:

\paragraph{A2539}
Caretta02 report that this cluster appears to be a superposition of
groups at $z \sim 0.175$ and $z \sim 0.186$. Our redshift estimate
falls between these two values, but is more consistent with the lower
value.  The rather high estimated velocity dispersion could be caused
by such a superposition of two groups along the line of sight,
although the asymmetry index and the normalized tail index do not
indicate a significant departure from a Gaussian distribution.

\paragraph{A2540} 
Our redshift value of $z = 0.134$ is significantly higher than the
value of $z = 0.129$ measured by Caretta02. However, our estimate is
based on a significantly larger number of redshifts.

\paragraph{A2546} 
This appears to form a double system with nearby \objectname{A2541},
which is located near the Southeastern end of our survey region. By
including additional literature redshifts of galaxies located slightly
outside our survey region, we measure a redshift
$z=0.1124^{+0.0014}_{-0.0011}$ and velocity dispersion of
$995^{+854}_{-423}\, {\rm km\, s}^{-1}$, based on 11 member galaxies.
This redshift value is consistent with the redshift reported by
Caretta02 ($z=0.1135$, based on 16 galaxies).

\paragraph{AQ2311.4-2126}
The coordinates of the cluster candidate \objectname{AqrCC~32}
identified by Caretta02 places it between our clusters
\objectname{AQ2311.4-2126} and \objectname{AQ2311.4-2134}, and it is
most likely a superposition of these two clusters.

\paragraph{A2550}
While this cluster was not recovered using our standard detection
scheme outlined in \S~\ref{sec:cluster}, a significant peak was
detected at this location when using all available literature
redshifts in the region. We note that the apparent center of this
cluster is located in a region where the completeness of our
spectroscopic survey is particularly low (see
Figs.~\ref{fig:BrightComplete} and~\ref{fig:FaintComplete}).  Our
measured redshift value $z=0.1151^{+0.0022}_{-0.0013}$, based on 14
galaxies, is discrepant from previously reported estimates: Caretta02
give a value $z = 0.1226$ from 6 galaxies, while
\citet{Kowalskietal1983} reported $z = 0.1543$ based on a single
redshift measurement. The latter value is close to the redshifts of
the clusters \objectname{AQ2311.4-2136} and \objectname{AQ2311.4-2134}
which are both located within $\sim 0.5\arcdeg$ from the ACO position
of \objectname{A2550}. On inspection of a deep $R$-band image of this
field, from our ESO WFI data (Dahle et~al.~2004; in preparation),
we find that the bright elliptical galaxy
\objectname{2MASX~J23113576-2144462} is dominating the center of
\objectname{A2550}. Its redshift was measured as $z = 0.1216$ by
\citet{Batuskietal1999}, similar to the cluster redshift value
reported by Caretta02. Furthermore, we note that line-of-sight
contamination from the very rich nearby cluster \objectname{A2554} may
have influenced our results, and the high velocity dispersion
($1427^{+410}_{-210}\, {\rm km\, s}^{-1}$) and asymmetry index (1.4)
of our galaxy sample indicates that we may be probing a filamentary
structure linked to \objectname{A2554} rather than \objectname{A2550}.

\paragraph{A2554}
Based on 23 cluster members, \citet{CollessHewett1987} measured a
velocity dispersion of $797^{+165}_{-114}$\, km\, s$^{-1}$, consistent
with our estimate at the $1\sigma$ level.

\paragraph{A2565}
Our redshift value is slightly discrepant from that reported by
Caretta02, but our cluster position is also significantly mis-aligned
with their position.  We note that this cluster is located in a region
where the spectroscopic completeness value is particularly low, and
our position is shifted in a direction towards a region where the
completeness is higher. Thus, we are probably sampling a slightly
different region than Caretta02, and the high asymmetry index
indicates that our redshift estimate may be significantly affected by
filaments or other structures outside the virial radius of the
cluster.

\paragraph{AQ2317.4-2223}
The coordinates of the cluster candidate \objectname{AqrCC~48},
described by Caretta02 as ``a superposition of small groups'', places
it between our clusters \objectname{AQ2317.4-2223} and
\objectname{AQ2318.9-2232}, and it is most likely a superposition of
these two objects.

\paragraph{A2576}
\citet{Batuskietal1995} reported a velocity dispersion measurement of
$630$\, km\, s$^{-1}$ for this cluster, which is somewhat lower than
the value we measure.  However, their measurement is based on only 10
member galaxies, so the uncertainty should be significantly higher
than for our measurement.

\paragraph{AQ2321.3-2224}
This cluster probably corresponds to the cluster candidate
\objectname{AqrCC~62} reported by Caretta02. While these authors did
not detect a concentration in redshift space at this location, we find
from our data a corresponding galaxy group (or poor cluster) at $z
\sim 0.193$.

\paragraph{A3996}
While the redshift measurement of Caretta02 ($z=0.0854$) is
significantly lower than our value, these authors report
\objectname{A3996} to be a superposition of small groups. Our redshift
value is quite similar to that reported by \citet{Batuskietal1999}
($z=0.1155$), based on 8 redshifts.  The high tail index and highly
uncertain velocity dispersion indicate that this cluster may be
significantly affected by foreground/background contamination.


\section{The Aquarius Superclusters}\label{sec:superclusters}

Based on a percolation analysis of Abell and ACO clusters in Aquarius,
\citet{Batuskietal1999} interpreted the region at $0.08 < z < 0.12$ as
a single supercluster structure. In contrast, Caretta02 found that the
Aquarius region contains two separate superclusters, seen in
projection, at $z\sim 0.08$ and $z \sim 0.11$. While
\citet{Batuskietal1999} argued that these two superclusters are
physically linked and form a single system, most of the evidence for
this link came from an erroneous redshift for \objectname{A2541}. In
fact, this cluster is situated within the densest portion of the most
distant of the two superclusters rather than in a region between the
superclusters. The cone plot of Fig.~\ref{fig:Cone1}, clearly shows
strong clustering at $z\sim 0.08$ and $z\sim 0.11$, separated by a
region of comparatively low density.

The overdensities and total masses of these structures can be
estimated from the measured galaxy velocity dispersions of their
member clusters. A lower limit on the supercluster mass can be found
by summing the estimated masses of all the clusters within their
respective virial radii, and an upper limit can be found by assuming a
given cluster density profile and integrating this for all clusters
over the volume covered by the supercluster.

We assume that the clusters generally follow the density profile of
cold dark matter halos proposed by \citet[][hereafter
NFW]{NavarroFrenkWhite1997}.  While this model has been shown to
provide a good fit to the radial cluster density profile in the
virialized region of clusters
\citep{Arabadjisetal2002,BivianoGirardi2003,Dahleetal2003b}, there are
currently only limited constraints on the density profile in the
infall region beyond the virial radius
\citep{Rinesetal2003,Kneibetal2003}. Also, clusters embedded in the
rare, high-density environment of a supercluster may have density
profiles that deviate significantly from those of average clusters.
With our weak lensing data set from WFI imaging, we will be able to
directly constrain the density profiles of clusters within the
superclusters and provide more accurate estimates of the supercluster
masses.  Our density estimates given below do not take into account
any redshift space distortions due to departures from the Hubble
flow. In reality, the cluster distribution could be either flattened
or elongated in the $z$ direction by cluster peculiar velocities.

For each cluster, we assume an NFW model with an effective velocity
dispersion $\sigma_v = \sqrt{50} H(z) r_{200}$, where $H(z)$ is the
Hubble parameter, and $r_{200}$ is the radius within which the average
mass density is 200 times the critical density at the relevant
redshift. The average mass within $r_{200}$ of the members of the
superclusters discussed below is $\langle M_{200} \rangle \sim 4
\times 10^{14} h^{-1} M_{\sun}$.  For this cluster mass, an NFW
cluster in a $\Lambda$CDM universe has a predicted concentration
parameter $C = 6$, which we will adopt as a general value.  In a
$\Lambda$CDM universe, the virial radius is actually slightly larger
than $r_{200}$, while the two are the same in an Einstein-de Sitter
universe.


\subsection{The $z\sim 0.11$ Supercluster}

As noted in \S~\ref{sec:population}, \citet{Batuskietal1999}
identified a particularly dense knot of five $R \ge 1$ ACO clusters
contained within a sphere of radius 10 $h^{-1}$ Mpc at $z=0.11$,
representing a spatial number density 150 times the mean density of $R
\ge 1$ ACO clusters.  By this measure, the Aquarius knot represents
the strongest overdensity of any known supercluster. The knot is very
obvious in Figure~\ref{fig:Cone1}, and corresponds to the densest
part of the supercluster.

All of the 5 ACO clusters forming the knot noted by
\citet{Batuskietal1999} are located within our survey field. In
\S~\ref{sec:velocity} (see the note regarding \objectname{A2546}) we
provide a new redshift estimate for another $R \ge 1$ ACO cluster,
\objectname{A2541}, which makes it a sixth member of the knot.  In
addition, we find that \objectname{AqrCC~24} plus two previously
unknown systems are also members of the knot at $z = 0.11$.

The lower limit on the supercluster mass obtained from the sum of the
$M_{200}$ values of the nine clusters within our survey region is $4.5
\times 10^{15} h^{-1} M_{\sun}$. Integrating NFW density profiles for
the nine galaxy clusters over the volume bounded by $23^{\rm h}
09^{\rm m} 00^{\rm s} < \alpha < 23^{\rm h} 22^{\rm m} 30^{\rm s}$,
$-23\degr 10 \arcmin < \delta < -21\degr 15 \arcmin$ and $0.109 < z <
0.114$, corresponding to a volume of $17 \times 8.5 \times 15 h^{-3}
{\rm Mpc}^3$, gives an estimate of the total mass of $8.3 \times
10^{16} h^{-1} M_{\sun}$.  The lower and upper limits for the mass
correspond to overdensities of $\rho = 17 \bar{\rho}$ and $\rho = 330
\bar{\rho}$ within this volume.


\subsection{The $z\sim 0.08$ supercluster}

We identify 11 clusters and groups in the most nearby Aquarius
supercluster at $0.084 < z < 0.091$, of which three are ACO clusters,
three more were detected by Caretta02 (\objectname{AqrCC~48}, which we
interpret to be a binary system, and \objectname{AqrCC~56}), and the
remaining five are new detections. The sum of their $M_{200}$ masses
is $3.2 \times 10^{15} h^{-1} M_{\sun}$, and their integrated masses
over a volume of $15 \times 7.9 \times 20 h^{-3} {\rm Mpc}^{3}$
(bounded by $23^{\rm h} 08^{\rm m} 00^{\rm s} < \alpha < 23^{\rm h}
22^{\rm m} 30^{\rm s}$, $-22\degr 55 \arcmin < \delta < -21\degr 15
\arcmin$ and $0.084 < z < 0.091$) is $8.3 \times 10^{16} h^{-1}
M_{\sun}$.  The lower and upper limits for the mass correspond to
overdensities of $\rho = 13 \bar{\rho}$ and $\rho = 330 \bar{\rho}$
within this volume.  We note that our Aquarius survey field is too
small to encompass the entire supercluster, and Caretta02 identify
several supercluster members outside our field.

The integrated masses of the $z=0.11$ and $z=0.08$ structures are
similar to the upper limit on the mass of the Corona Borealis
supercluster ($M < 8 \times 10^{16} h^{-1} M_{\sun}$) estimated by
\citet{Smalletal1998}.


\section{Summary}\label{sec:summary}

In this paper we have presented a spectroscopic analysis of $\sim4000$
galaxies down to $R<19.5$ in the direction of Aquarius and Cetus.
These observations form part of our wider efforts to conduct a
spectro-photometric and weak lensing survey of these two regions.  The
spectroscopic data are sensitive to structures out to $\sim$1~Gpc or
equivalently $z\sim0.4$. These data will greatly aid our ongoing
efforts to measure the dark matter distribution in these regions and
quantify the bias over a large range of scales and environments.

We described the survey that we are conducting, focusing on the
spectroscopic input catalogue. We then presented our 2dF observations
and the reduction of the spectra. We characterized the spectroscopic
sample by considering the median spectra, finding evolution of the
median galaxy spectrum with redshift. This evolution was attributed to
selection effects, rather than any true evolution in the galaxy
population. We presented spatial diagrams for the two target regions
and also quantified the number redshift distributions.  These
displayed a large amount of structure and were poorly fit by a
2dFGRS-like number redshift distribution.  For our future lensing
analysis, where these quantities are required, we will use direct fits
to the data. Using the full redshift space sample we then identified
clusters in both the Aquarius and Cetus regions. In Aquarius we found
a total of 48 clusters and galaxy groups, 26 of which had not before
been noted, and for 5 of those that had, we provided the first
spectroscopic redshift estimates.  For Cetus we found 14 clusters and
galaxy groups, of which 12 were previously unknown systems. For each
candidate system we then provided estimates of the centroid redshift
and velocity dispersion.  We showed that there are two superclusters
in Aquarius, one at redshift $z=0.08$ and the other at $z=0.11$. For
these we estimated the overdensities and total masses, finding that
they were similar to the Corona Borealis supercluster.

In the future, we intend to characterize the galaxies through their
spectral energy distributions using a principal components analysis
\citep{Madgwicketal2002}. This will allow us to explore the putative
result that early-type galaxy light is an unbiased tracer of the dark
matter \citep{Kaiseretal1998,Grayetal2002,Dahleetal2002}.  In addition
to the primary goals, we anticipate that when fully complete this
survey will provide other valuable scientific information. For
example, comparison of the weak lensing data with the optically
identified clusters and with the archival X-ray data, will allow us to
assess the objectivity of optical and X-ray luminosity selection
methods for cluster surveys \citep{Dahleetal2003}.  This will be
important to quantify if cluster abundance surveys that aim to
constrain cosmological parameters are to be taken seriously in this
new age of precision.


\section*{Acknowledgments}

RES acknowledges a PPARC postdoctoral research assistantship.  HD and
PBL acknowledge support from The Research Council of Norway. We thank:
Bryn Jones for providing useful image extraction software; Will
Sutherland and the 2dFGRS team for providing the redshift estimation
code {\ttfamily CRCOR}; Timothy C. Beers for providing his {\ttfamily
  ROSTAT} code and advice on its use; the support staff at the AAT.
This research has made use of: the SuperCOSMOS archives at the Royal
Observatory Edinburgh; the NASA/IPAC Extragalactic Database, at the
Jet Propulsion Laboratory.



\begin{thebibliography}{}
  
\bibitem[{Abell}(1961){1961}]{Abell1961} {Abell}, G.~O.\ 1961, \aj,
  66, 607
  
\bibitem[{Abell}, {Corwin} \& {Olowin}(1989){Abell}, {Corwin} and
  {Olowin}]{Abelletal1989} {Abell}, G., {Corwin}, H., \& {Olowin}, R.\ 
  1989, \apjs, 70, 1; (ACO)
  
\bibitem[Arabadjis, Bautz, \& Garmire(2002)]{Arabadjisetal2002}
  Arabadjis, J.~S., Bautz, M.~W., \& Garmire, G.~P.\ 2002, \apj, 572,
  66
  
\bibitem[{Bacon} \& {Taylor}(2003){Bacon} and
  {Taylor}]{BaconTaylor2003} {Bacon}, D., \& {Taylor}, A.~N., 2003,
  MNRAS, 344, 1307
  
\bibitem[{Bailey}, {Glazebrook} \& {Bridges}(2002){Bailey},
  {Glazebrook} and {Bridges}] {Baileyetal2002} {Bailey}, J.,
  {Glazebrook}, K., \& {Bridges}, Y.\ 2002, ``The 2dF user manual'',
  {\ttfamily http://www.aao.gov.au/2df/manual.html}; (BGB02)
  
\bibitem[{Batuski} {et~al.}(1995){Batuski}]{Batuskietal1995}
  {Batuski}, D.~J., {Maurogordato}, S., {Balkowski}, C., \& Olowin,
  R.~P.\ 1995, \aap, 294, 677
  
\bibitem[{Batuski} {et~al.}(1999){Batuski}]{Batuskietal1999}
  {Batuski}, D.~J., {Miller}, C.~J., {Slinglend}, K.~A., {Balkowski},
  C., {Maurogordato}, S., {Cayatte}, V., {Felenbok}, P., \& {Olowin},
  R.\ 1999, \apj, 520, 491
  
\bibitem[{Beers} {et~al.}(1990){Beers}]{Beersetal1990} {Beers}, T.~C.,
  {Flynn}, K., \& {Gebhardt}, K.\ 1990, \aj, 100, 32
  
\bibitem[{Benson} {et al.}(2000){Benson}]{Bensonetal2000} Benson, A.
  J., Cole, S., Frenk, C. S., Baugh, C. M., \& Lacey, C. G., \ 2000,
  MNRAS, 311, 793
  
\bibitem[{Bird} \& {Beers}(1993){Bird} and {Beers}]{BirdBeers1993}
  {Bird}, C.~M., \& {Beers}, T.~C.\ 1993, \aj, 105, 1596
  
\bibitem[Biviano \& Girardi(2003)]{BivianoGirardi2003} Biviano, A., \&
  Girardi, M.\ 2003, \apj, 585, 205
  
\bibitem[{Blanton} {et~al.}(2003){Blanton}]{Blantonetal2003}
  {Blanton}, M., \& The SDSS Team,\ 2003, \apj, 592, 819
  
\bibitem[{Borgani} {et~al.}(1999){Borgani}]{Borganietal1999}
  {Borgani}, S., {Girardi}, M., {Carlberg}, R.~G., {Yee}, H.~K.~C., \&
  {Ellingson}, E.\ 1999, \apj, 527, 572
  
\bibitem[{Caretta} {et~al.}(2002){Caretta}]{Carettaetal2002}
  {Caretta}, C., {Maia}, M., {Kawasaki}, W., \& {Willmer}, C.\ 2002,
  \aj, 123, 1200; (Caretta02)
  
\bibitem[{Ciardullo} {et~al.}(1985){Ciardullo}]{Ciardulloetal1985}
  {Ciardullo}, R., {Ford}, H., \& {Harms}, R.\ 1985, \apj, 293, 69
  
\bibitem[{Coleman}, {Wu} \& {Weedman}(1980){Coleman}, {Wu} and
  {Weedman}]{Colemanetal1980} {Coleman}, G. D., {Wu}, C.-C., \&
  {Weedman}, D. W.\ 1980, \apjs, 43, 393
  
\bibitem[{Colless} \& {Hewett}(1987){Colless} and
  {Hewett}]{CollessHewett1987} {Colless}, M., \& {Hewett}, P.\ 1987,
  \mnras, 224, 453
  
\bibitem[{Colless} {et~al.}(2001)]{Collessetal2001} {Colless}, M., \&
  The 2dFGRS Team,\ 2001, \mnras, 328, 1039
  
\bibitem[{Dahle} {et~al.}(2002){Dahle}]{Dahleetal2002} {Dahle}, H.,
  {Kaiser}, N., {Irgens}, R., {Lilje}, P.~B., \& {Maddox}, S.\ 2002,
  \apjs, 139, 313
  
\bibitem[{Dahle} {et~al.}(2003a){Dahle}]{Dahleetal2003} {Dahle}, H.,
  {Pedersen}, K., {Lilje}, P.~B., {Maddox}, S.~J., \& {Kaiser}, N.  \ 
  2003a, \apj, 591, 662

\bibitem[Dahle, Hannestad, \& Sommer-Larsen(2003b)]{Dahleetal2003b}
  Dahle, H., Hannestad, S., \& Sommer-Larsen, J.\ 2003b, \apjl, 588,
  L73
  
\bibitem[{Efron}(1987){Efron}]{Efron1987} {Efron}, B.\ 1987, Journal
  of the American Statistical Association, 82, 171

\bibitem[{Efstathiou} \& {Moody}(2001){Efstathiou} and
  {Moody}]{EfstathiouMoody2001} {Efstathiou}, G., \& {Moody}, S.\ 
  2001, \mnras, 325, 1603

\bibitem[{Girardi} {et~al.}(1993){Girardi}]{Girardietal1993}
  {Girardi}, M., {Biviano}, A., {Giuricin}, G., {Mardirossian}, F., \&
  {Mezzetti}, M.\ 1993, \apj, 404, 38

\bibitem[{Gray} {et~al.}(2002){Gray}]{Grayetal2002} {Gray}, M.~E.,
  {Taylor}, A.~N., {Meisenheimer}, K., {Dye}, S., {Wolf}, C., \&
  {Thommes}, E.\ 2002, \mnras, 568, 141

\bibitem[{Hambly} {et~al.}(2001a){Hambly}]{Hamblyetal2001a} {Hambly},
  N.~C., {MacGillivray}, H.~T., {Read}, M.~A., {Tritton}, S.~B.,
  {Thomson}, E.~B., {Kelly}, B.~D., {Morgan}, D.~H., {Smith}, R.~E.,
  {Driver}, S.~P., {Williamson}, J., {Parker}, Q.~A., {Hawkins},
  M.~R.~S., {Williams}, P.~M., \& {Lawrence}, A.\ 2001a, \mnras, 326,
  1279

\bibitem[{Hambly} {et~al.}(2001b){Hambly}]{Hamblyetal2001b} {Hambly},
  N., {Irwin}, M.~J., \& {MacGillivray}, H.~T.\ 2001b, \mnras, 326,
  1315

\bibitem[{Hoekstra} {et~al.}(2001){Hoekstra}]{Hoekstraetal2001}
  {Hoekstra}, H., {Yee}, H.~K.~C., \& {Gladders}, M.~D.\ 2001, \apjl,
  558, L11

\bibitem[{Hoekstra} {et~al.}(2002){Hoekstra}]{Hoekstraetal2002}
  {Hoekstra}, H., {van Waerbeke}, L., {Gladders}, M.~D., {Mellier},
  Y., \& {Yee}, H.~K.~C.\ 2001, \apjl, 558, L11

\bibitem[{Irgens} {et~al.}(2002){Irgens}]{Irgensetal2002} {Irgens},
  R.~J., {Lilje}, P.~B., {Dahle}, H., \& {Maddox}, S.~J.\ 2002, \apj,
  579, 227

\bibitem[{Kaiser}(1992){Kaiser}]{Kaiser1992} {Kaiser}, N.\ 1992. \apj,
  388, 272

\bibitem[{Kaiser} \& {Squires}(1993){Kaiser} and
  {Squires}]{KaiserSquires1993} {Kaiser}, N., \& {Squires}, G.\ 1993.
  \apj, 404, 441

\bibitem[{Kaiser}, {Squires} \& {Broadhurst}(1995)]{Kaiseretal1995}
  {Kaiser}, N., {Squires}, G., {Broadhurst}, T.\ 1995, \apj, 449, 460
  
\bibitem[{Kaiser} {et~al.}(1998){Kaiser}]{Kaiseretal1998} {Kaiser},
  N., {Wilson}, G., {Luppino}, G., {Kofman}, L., {Gioia}, I.,
  {Metzger}, M., \& {Dahle}, H.\ 1998, astro-ph/9809268

\bibitem[Kneib et al.(2003)]{Kneibetal2003} {Kneib}, J., {Hudelot},
  P., {Ellis}, R.~S., {Treu}, T., {Smith}, G.~P., {Marshall}, P.,
  {Czoske}, O., {Smail}, I., \& {Natarajan}, P.  \ 2003, \apj, 598,
  804

\bibitem[{Kowalski}, {Ulmer} \& {Cruddace}(1983)]{Kowalskietal1983}
  {Kowalski}, M.~P., {Ulmer}, M.~P., \& {Cruddace}, R.~G.\ 1983, \apj,
  268, 540

\bibitem[{Madgwick} {et~al.}(2002){Madgwick}]{Madgwicketal2002}
  {Madgwick}, D., \& The 2dFGRS Team, \ 2002, \mnras, 333, 133

\bibitem[{Murray} {et~al.}(1978){Murray}]{Murrayetal1978} {Murray},
  S.~S., {Forman}, W., {Jones}, C., \& {Giacconi}, R.\ 1978, \apjl,
  219, L89

\bibitem[{Navarro}, {Frenk}, \& {White}(1997)]{NavarroFrenkWhite1997}
  Navarro, J.~F., Frenk, C.~S., \& White, S.~D.~M.\ 1997, \apj, 490,
  493; (NFW)

\bibitem[{Peacock} \& {Smith}(2000){Peacock} and
  {Smith}]{PeacockSmith2000} Peacock, J.~A., Smith, R.~E., \ 2000,
  MNRAS, 318, 1144

\bibitem[{Percival} {et~al.}(2004){Percival}]{Percivaletal2004}
  Percival, W.~J., \& The 2dFGRS Team,\ 2004, MNRAS, submitted
  (astro-ph/0406513)
  
\bibitem[{Postman}, {Geller} \& {Huchra}(1986){Postman}, {Geller} and
  {Huchra}]{Postmanetal1986} {Postman}, M., {Geller}, M., \& {Huchra},
  J.\ 1986, \aj, 92, 1238
  
\bibitem[{Raychaudhury}(1989){Raychaudhury}]{Raychaudhury1989}
  {Raychaudhury}, S.\ 1989, \nat, 342, 251

\bibitem[{Rines} {et~al.}(2003)]{Rinesetal2003} Rines, K., Geller,
  M.~J., Kurtz, M.~J., \& Diaferio, A.\ 2003, \aj, 126, 2152

\bibitem[{Romer} {et~al.}(2000)]{Romeretal2000} {Romer}, A.~K.,
  {Nichol}, R.~C., {Holden}, B.~P., {Ulmer}, M.~P., {Pildis}, R.~A.,
  {Merrelli}, A.~J., {Adami}, C., {Burke}, D.~J., {Collins}, C.~A.,
  {Metevier}, A.~J., {Kron}, R.~G., \& {Commons}, K.\ 2000, \apjs,
  126, 209

\bibitem[{Schlegel}, {Finkbeiner} \& {Davis}(1998){Schlegel},
  {Finkbeiner} and {Davis}] {Schlegeletal1998} {Schlegel}, D.~J.,
  {Finkbeiner}, D.~P., \& Davis, M.\ 1989, \apj, 500, 525

\bibitem[{Schneider}(1998){Schneider}]{Schneider1998} {Schneider}, P.\ 
  1998, \apj, 498, 43

\bibitem[{Small} {et~al.}(1998){Small}]{Smalletal1998} Small, T.~A.,
  Ma, C., Sargent, W.~L.~W., \& Hamilton, D.\ 1998, \apj, 492, 45

\bibitem[{Strauss} {et~al.}(2002){Strauss}]{Straussetal2002}
  {Strauss}, M.~A., \& The SDSS Team, \ 2002, \aj, 124, 1810
  
\bibitem[{Taylor} {et~al.}(2004){Taylor}]{Tayloretal2004} Taylor,
  A.~N., Bacon, D.~J., Gray, M.~E., Wolf, C., Meisenheimer, K., Dye,
  S., Borch, A., Kleinheinrich, M., Kovacs, Z., \& Wisotzki, L., 2004,
  MNRAS, submitted (astro-ph/0402095)

\bibitem[{Tegmark} {et~al.}(2004){Tegmark}]{Tegmarketal2004} Tegmark,
  M., \& The SDSS Team, \ 2004, \apj, 606, 702

\bibitem[{Tonry} \& {Davis}(1979){Tonry} and {Davis}]{TonryDavis1979}
  {Tonry}, J., \& {Davis} M.\ 1979, \aj, 84, 1511

\bibitem[{Wang} {et~al.}(2000){Wang}]{Wangetal2000} Wang, L.,
  Caldwell, R.~R., Ostriker, J.~P., \& Steinhardt, P.~J.\ 2000, \apj,
  530, 17
  
\bibitem[{Willis}, {Hewett} \& {Warren}(2001){Willis}, {Hewett} and
  {Warren}]{Willisetal2001} {Willis}, J., {Hewett}, P., \& {Warren},
  S.\ 2001, \mnras, 325, 1002

\end{thebibliography}
\end{document}